# 3D RNA and functional interactions from evolutionary couplings

Caleb Weinreb[1], Adam Riesselman[1], John B. Ingraham[1], Torsten Gross[1,2], Chris Sander[3], Debora S. Marks[1]

[1] Department of Systems Biology, Harvard Medical School, Boston, MA 02115, USA
[2] Institute of Pathology, Charité – Universitätsmedizin Berlin, 10117 Berlin, Germany
[3] Department of Cell Biology, Harvard Medical School, Boston, MA 02115, USA

Correspondence: debbie@hms.harvard.edu

## Summary

Non-coding RNAs are ubiquitous, but the discovery of new RNA gene sequences far outpaces research on their structure and functional interactions. We mine the evolutionary sequence record to derive precise information about function and structure of RNAs and RNA-protein complexes. As in protein structure prediction, we use maximum entropy global probability models of sequence co-variation to infer evolutionarily constrained nucleotide-nucleotide interactions within RNA molecules, and nucleotide-amino acid interactions in RNA-protein complexes. The predicted contacts allow all-atom blinded 3D structure prediction at good accuracy for several known RNA structures and RNA-protein complexes. For unknown structures, we predict contacts in 160 non-coding RNA families. Beyond 3D structure prediction, evolutionary couplings help identify important functional interactions, e.g., at switch points in riboswitches and at a complex nucleation site in HIV. Aided by accelerating sequence accumulation, evolutionary coupling analysis can accelerate the discovery of functional interactions and 3D structures involving RNA.

## Introduction

RNAs have diverse known biological roles (Garneau et al., 2007; Huang et al., 2015; Martin and Ephrussi, 2009; McManus and Graveley, 2011; Olsen et al., 1990; Rutherford et al., 2015; Sigova et al., 2015; Warf and Berglund, 2010) and both genetic and biochemical screens suggest reservoirs of functional RNA molecules we know little about. For instance, transcriptional profiling has revealed large numbers of non-coding RNA genes, with unknown functions beyond context specific expression (Eddy, 2014; Rinn and Chang, 2012). Although much of this transcription may be biological noise, these screens have revealed some novel long non-coding RNAs that may have three-dimensional structures and can act – for example – as protein scaffolds (Quinodoz and Guttman, 2014). High-throughput biochemical screens, adapted from earlier work on RNA foot-printing (Ehresmann et al., 1987; Latham and Cech, 1989; Moazed and Noller, 1986), have identified transcriptome-wide RNA base-pairing *in vivo* (Ding et al., 2014; Rouskin et al., 2014; Spitale et al., 2015; Wan et al., 2014) including within the coding region of mRNAs, suggesting a function independent of coding potential.

Many of these newly observed RNAs may have specific 3D structures (Mortimer et al., 2014; Novikova et al., 2012). Since high-resolution structure determination remains labor intensive, there is a renewed interest in computational prediction of RNA 3D structure and identification of functional interactions. One long-standing approach for inferring RNA structure is to search for pairs of positions that show correlated substitutions in alignments of homologous sequences. This approach was used to define the 1969 Levitt model of tRNA (Levitt, 1969), the Fox-Woese model of 5S rRNA (Fox and Woese, 1975) and the Michel-Westhof model of a group 1 ribozyme (Michel and Westhof, 1990). Comparative sequence analyses of RNA continue to contribute to successful RNA sequence alignment (Nawrocki and Eddy, 2013) and secondary structure prediction methods (Hofacker et al., 2002; Nussinov and Jacobson, 1980; Rivas and Eddy, 1999; Zuker, 2003) (review (Hofacker and Lorenz, 2014)), but existing techniques for identifying correlated positions have been less successful at detecting key tertiary contacts that are not involved in Watson-Crick base pairing (Dutheil et al., 2010).

The inability of existing methods to detect long-range tertiary contacts from sequence covariation has limited the progress of purely *in silico* RNA 3D structure prediction, since RNAs with multiple helical segments can assume diverse folds, producing a conformational space that is impossible to search effectively. Thus, despite rapid strides in 3D structure prediction accuracy for small (< 40 nt) RNAs (Cao and Chen, 2011; Das and Baker, 2007; Das et al., 2010; Frellsen et al., 2009; Parisien and Major, 2008), predicting the structure of large (> 70 nt) RNAs remains challenging (Laing and Schlick, 2010; Miao et al., 2015), unless experimental restraints are available from biochemical probing data (Cheng et al., 2015; Magnus et al., 2014; Ramani et al., 2015).

Why have existing methods that compute pairwise patterns of sequence co-variation in RNA (Mokdad and Frankel, 2008 ; Pang et al., 2005; Shang et al., 2012) had limited success at detecting tertiary interactions? One possibility is that RNA tertiary contacts often form complex networks (Butcher and Pyle, 2011) in which patterns of sequence constraints can interfere with each other, obscuring true interactions and producing spurious transitive correlations when multiple contacts are chained together. A similar problem stymied protein structure prediction until the application of global maximum entropy models that could de-convolve the underlying network of constrained residue-residue interactions (Hopf et al., 2012; Hopf et al., 2014; Marks et al., 2011; Marks et al., 2012; Morcos et al., 2011; Ovchinnikov et al., 2014; Weigt et al., 2009).

Here we adapted the maximum entropy model to RNA sequence alignments and tested the ability of the approach to predict tertiary structure contacts on 180 RNA gene families representing thousands of RNA genes. Comparisons to known structures confirm the accuracy of contact prediction and 3D folding results. We further extend the model to protein-RNA interactions with accurate prediction of 6 RNA-protein interactions.

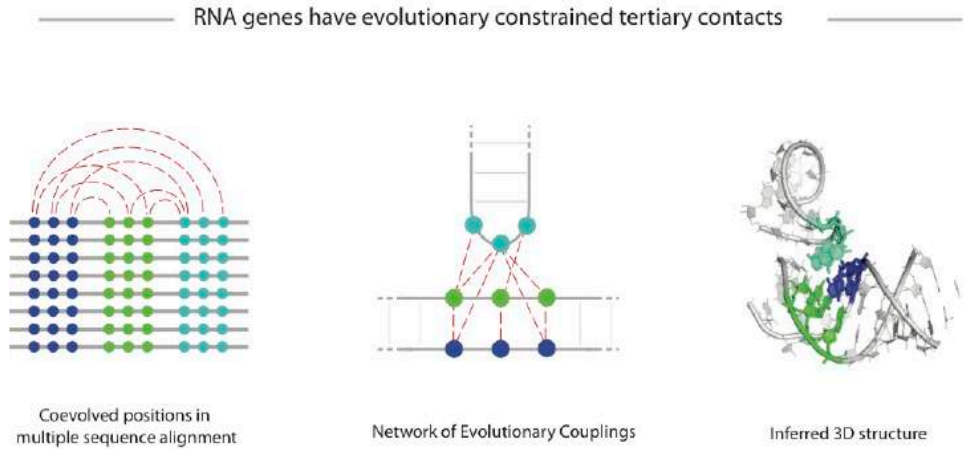

**Figure 1: Inferring RNA structure from sequence coevolution**
Non-coding RNAs form 3D structures stabilized by complex networks of secondary and tertiary interactions. In many cases, these interactions leave an evolutionary imprint reflecting epistasis between contacting nucleotides. Computationally detecting these interactions in multiple sequence alignments can reveal RNA 3D structure.

## Results

### Evolutionary couplings (ECs) accurately predict 3D contacts

We adapted the evolutionary couplings model – which we previously used to predict contacts within proteins – to calculate ECs for RNA (Figure 1). Briefly, we model each RNA family as distribution over sequence space, where the probability of a sequence $\sigma$ reflects the single-site biases $h_i$ at each position $i$, and the coupling terms $J_{ij}$ between each pair of positions $(i, j)$, as follows.

$$P(\boldsymbol{\sigma}) = \frac{1}{Z}\exp\left(\sum_{i=1}^{L} h_i(\sigma_i) + \sum_{i=1}^{L-1}\sum_{j=i+1}^{L} J_{ij}(\sigma_i, \sigma_j)\right).$$

We fit this model to natural sequence distributions using an approach based on penalized Maximum Likelihood estimation. We refer to pairs of positions with the strongest coupling terms as "evolutionary couplings." See Methods for details.

Evolutionary couplings (ECs) between pairs of nucleotides were evaluated by their ability to recover RNA 3D contacts in the known structure of a representative sequence from each of 22 RFAM families (Data S1). To assess the accuracy of predicted contacts, we define contacts as true positives if their minimum-atom-distance is < 8 Å. All 22 RNAs with known structures had a true positive rate above 70% for the top ranked L/2 ECs (where L is the number of nucleotides). ECs predicted contacts with greater accuracy than mutual information (MI), which has been widely used for RNA secondary structure prediction (Freyhult et al., 2005; Gutell et al., 1992) (Figure 2A). This held even when using enhanced MI that implements two features of the EC statistical model: (1) Down-weighting of similar sequences to avoid spurious correlations from phylogeny; (2) An average product correction (APC) (Dunn et al., 2008). We refer to MI without these modifications as raw MI ($MI_R$), and denote the enhanced MI as $MI_E$.

There are two reasons why the global model for ECs may perform better than MI. First, nucleotides that are strongly conserved will not display high mutual information, but may still have a high EC score. Second, false positive transitive correlations score highly with local methods such as MI as each pair is computed without reference to the whole network of interactions, whereas ECs successfully de-convolve transitive correlations.

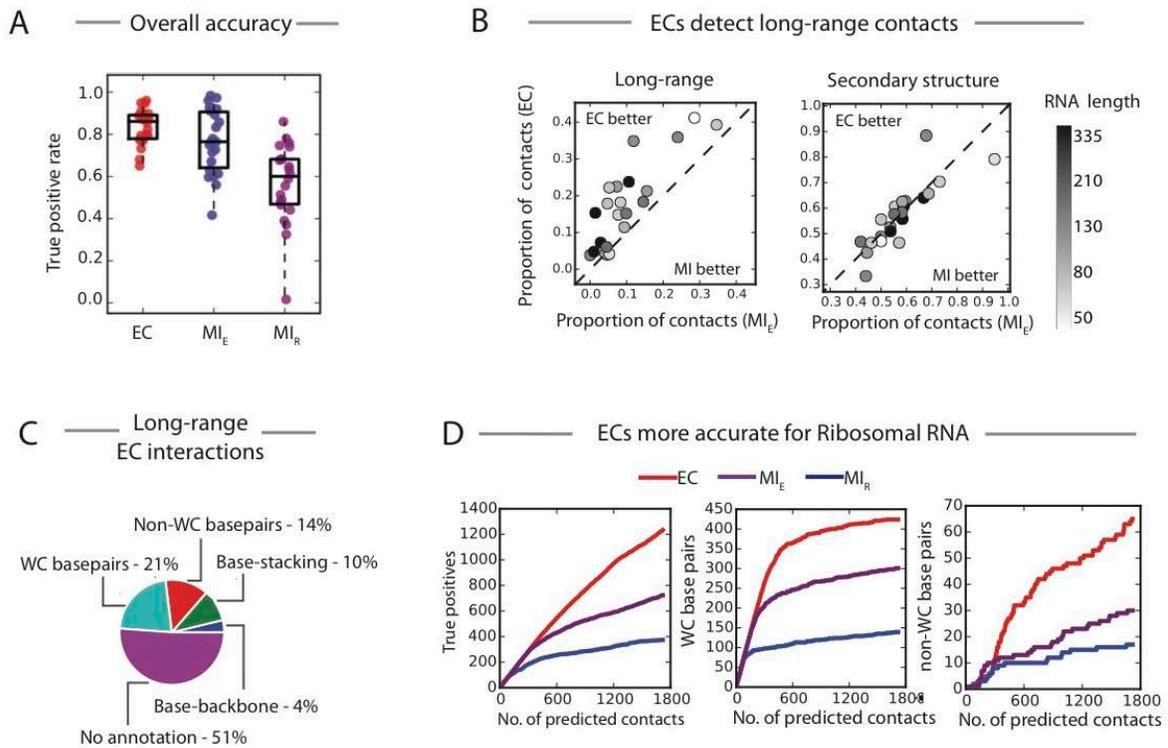

**Figure 2: Comparison of EC to MI: Summary of 22 RNA families**
(A) Evolutionary couplings (ECs) predict 3D contacts with a higher overall accuracy than $MI_E$ or $MI_R$ on a test set of 22 RNA families (Data S1). (B) Whereas EC and $MI_E$ detect a similar number of secondary structure contacts, ECs are significantly enriched with long-range contacts. (C) EC-inferred long-range contacts represent a variety of biochemical interactions annotated from the crystal structure. (D) ECs for the 40S ribosome (RF01960) are dramatically more accurate than MI, detecting more true positives overall (left), more Watson-Crick base pairs (middle) and more non-WC base pairs (right). ECs also detect ~50 long-range contacts that bridge distance parts of the secondary structure (Figure S2).

### Evolutionary couplings detect long-range contacts and non Watson-Crick base pairs

Though overall accuracy is important, not all contacts are 'created equal'. Often, complex RNA folds are stabilized by a small number of critical long-range contacts that bridge distant parts of the secondary structure. Though ECs of long-range contacts (definition in Methods) have lower scores than those in secondary structures (Figure S1), we nevertheless detect long-range contacts robustly across 22 RFAM families, with an average of $0.07 * L$ long-range contacts among the top L/2 ECs for an RNA of length $L$. This represents a substantial improvement over previous methods, since ECs contained 2.4 times more long-range contacts than $MI_E$, (0.8 – 10.8 times more across 22 individual examples; $p \leq 10^{-5}$ using a paired t-test; Figure 2B).

Pairs of RNA bases often form contacts through hydrogen bonding, assuming geometrical configurations that can broadly be divided into Watson-Crick (WC) base pairs and non-WC base pairs. Though covariation has long been used to infer Watson-Crick (WC) base pairs, the strength of coevolution for non-WC base pairs is an area of open investigation (Dutheil et al., 2010). We found that ECs are sensitive to non-WC base pairs, with the top L/2 ECs containing 16% of all annotated non-WC base pairs across the 22 structures – 1.7 times more on average than MI (0.5 – 4.0 times more across 22 individual examples; $p \leq 0.002$ using a paired t-test). We expect that ECs will complement existing approaches for detecting non-WC base pairs that rely on the concept of isostericity (Lescoute et al., 2005; Mokdad and Frankel, 2008) since they can focus attention on interactions with the strongest co-evolutionary signal.

### Evolutionary couplings reveal contacts in the eukaryotic ribosome

ECs may be sensitive to interacting nucleotides in large RNAs that form topologically complex folds with abundant long-range contacts. ECs computed on a full alignment of eukaryotic ribosomal sequences (RF01960) are over 90% accurate for the top 900 (L/2) contacts and predict substantially more WC and non-WC base pairs than $MI_E$ (Figure 2D), as well as more long-range contacts between nucleotides (Figure S2). In addition to increased sensitivity, ECs also have greater specificity than $MI_E$, with 2.8-fold fewer false-positive predictions. Capturing complex networks of contacts in large RNAs will be crucial for decoding the structure of lncRNAs and structured viral genomes.

### Long-range contacts allow accurate 3D structure prediction

We hypothesized that EC-derived long-range contacts could benefit 3D structure prediction for medium-sized RNAs, which has so far not been possible with secondary structure alone. Using coarse-grained molecular dynamics implemented in NAST (Butcher and Pyle, 2011; Jonikas et al., 2009) followed by simulated annealing with XPLOR (Schwieters et al., 2003), we predicted candidate all-atom structures (Figure 3A) for representatives from five selected RNA families (selection based on length, 70-120 nts and at least one *highly-long-range* contact, see Methods). In each case, the best of four candidate predictions had an all atom RMSD of 7 - 10 Å, comparable to the state of the art for RNA structure predictions that use tertiary contacts derived from biochemical probing (6.8 – 11.7 Å) (Miao et al., 2015) and a dramatic improvement over pure *in silico* structure prediction, where the average RMSD is 20 Å for medium size RNAs (50 – 130 nt) (Laing and Schlick, 2010). Our method shows a high level of precision, with 16/20 (4 predicted structures each for 5 RNA families = 20 total) predicted structures having correct orientation of the helices (Figure S3), and all predictions significantly closer to the experimental structure than controls folded without tertiary constraints (Figure 3B, p < 0.01 for all five comparisons; see Data S2 for RMSD and quality structure metrics; see Figure S4 for plots of RMSD vs. NAST energy).

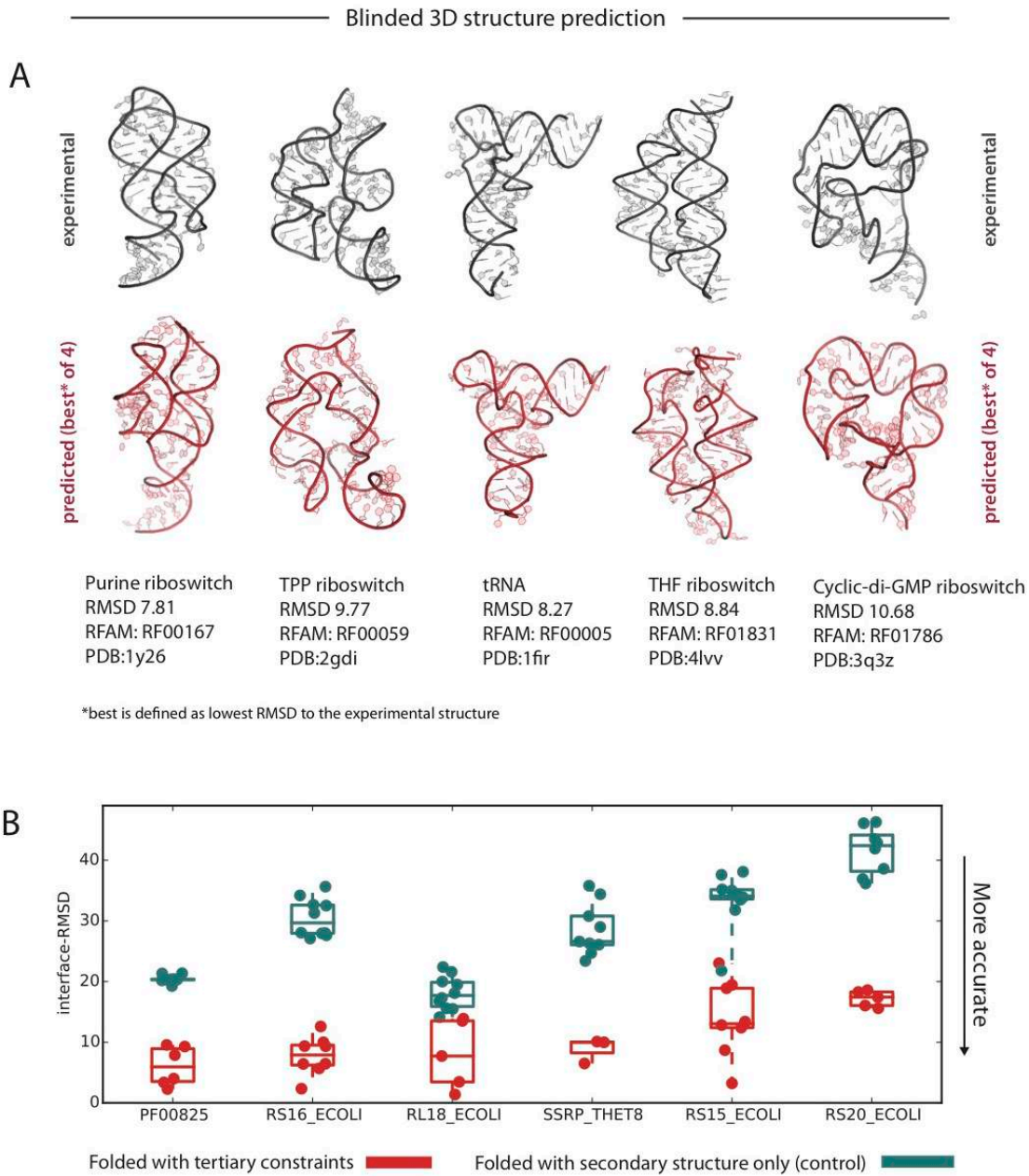

**Figure 3: Evolutionary couplings significantly improve 3D structure prediction accuracy**
We predicted all-atom 3D structures for five RNA families using evolutionary couplings as distance restraints (four candidate structures per family). (A) The candidate (red) with lowest RMSD to the experimental structure (gray) is shown for each family. (B) We performed folding controls with secondary structure only and found that they had significantly higher deviation from the experimental structure than models folded with EC-derived tertiary contacts. For full results on all predicted models, see Figures S3,S4 and Data S2,S7.

## Evolutionary couplings identify intermolecular contacts in RNA-protein complexes

Since ECs can detect contacts in RNA and protein separately, we next investigated whether they can reveal intermolecular contacts in 21 RNA-protein complexes with known structure (Data S3, see Methods for selection criteria), including 19 ribosomal proteins bound to the bacterial 16S rRNA and the ribonucleoprotein (RNP) complexes RNaseP and tmRNA (Figure 4). ECs predict RNA-protein contacts with high accuracy, as long as there is sufficient sequence diversity (Figure S5A). If contact predictions with ≥ 75% true positives for the top 4 contacts are defined as "highly accurate," then only 3/18 complexes with less than one effective sequence per nucleotide ($M_{eff}/L < 1$) had highly accurate predictions, whereas all (3/3) complexes with ($M_{eff}/L > 1$) had highly accurate predictions. One striking example is the 5S ribosomal protein RL18 – with ~ 4 effective sequence per nucleotide – which had 80% true positives for the top 10 inter-molecular contacts.

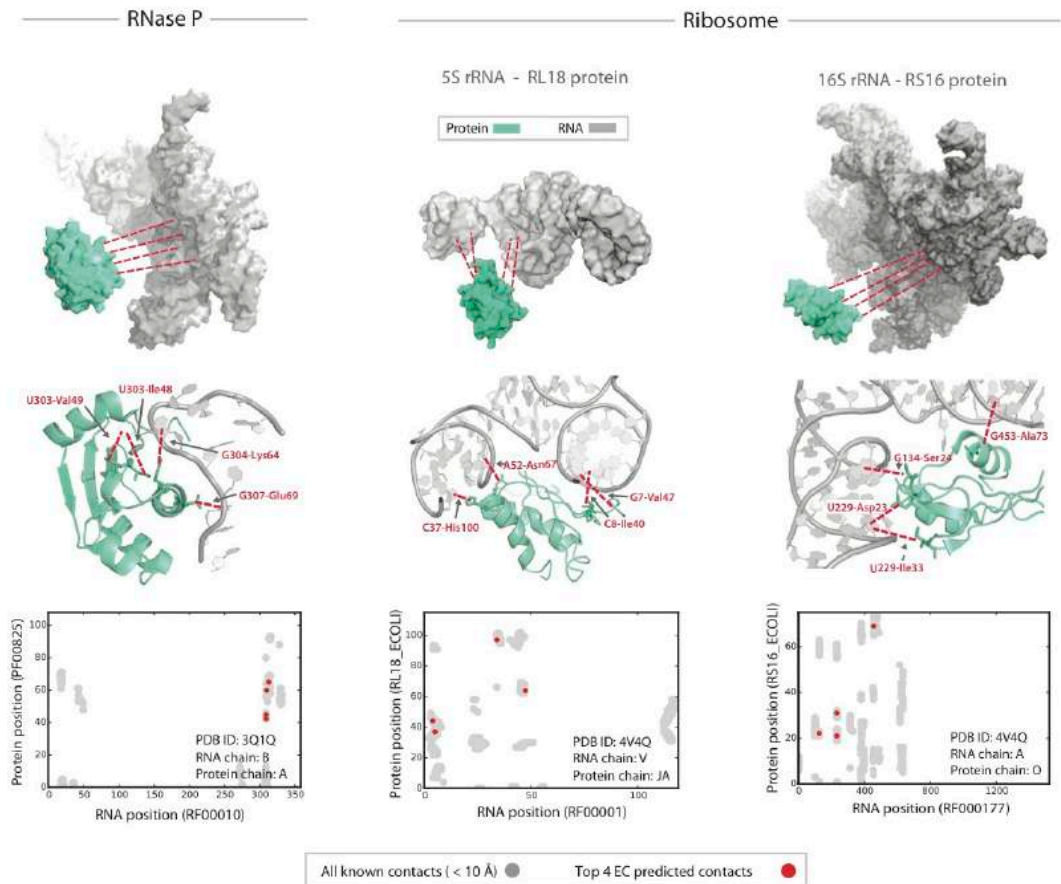

**Figure 4: Evolutionary couplings detect RNA-protein interactions**
Functionally important interactions between RNA nucleotides and protein amino acids may be constrained in evolution and detectable in multiple sequence alignments. We calculated ECs from phased alignments of 21 RNA-protein complexes, with three examples shown here (ribonucleoportein complex RNase P on the left and two proteins from the bacterial ribosome on the right, chosen as illustrative examples). In each case, we plot the top 4 highest-ranking ECs in cartoon style (top), super-imposed on the true structure (middle, in PDB numbering) and as red dots in a contact map (bottom). These interactions anchor their respective RNA-protein interfaces at multiple points of contact and open the door to sequence-based 3D structural studies of RNA protein complexes. For related results, see Figure S5 and Data S4,S5. For predicted PDB structures and phased alignments, see Data S7.

The accuracy of evolutionary couplings provides an exciting opportunity to predict the structure of RNA-protein complexes from sequence alone. We used EC-derived contacts as distance restraints for rigid body docking in HADDOCK (Dominguez et al., 2003), focusing on the 6/21 complexes with ≥ 75% true positives for the top 4 predicted contacts. In all 6 cases, structures docked with ECs had significantly lower i-RMSD than control structures docked with center of mass constraints only ($p < 10^{-4}$ for all 6 comparisons of mean i-RMSD between cases and controls, see Data S4 and Figure S5B). In 4/6 cases, one of the top three predicted structures had i-RMSD < 3.5Å, whereas the minimum i-RMSD among all controls (docked without EC distance restraints) was 15Å.

### Evolutionary Couplings highlight functional interactions

Among all instances of proximity in structured RNAs, only a small fraction is critical for function. We next asked whether evolutionary couplings enrich for these functionally important interactions by investigating the top scoring ECs in riboswitches, which are cis-acting regulatory segments of mRNAs that undergo ligand or temperature dependent conformational changes between at least two mutually exclusive functional states (Garst et al., 2011; Serganov and Patel, 2012). In four riboswitches from our dataset – *S*-adenosylmethionine (SAM), active vitamin $B_1$ (TPP), active folate vitamin (THF), and Adenine-sensing – we found a cluster of tertiary contacts that stabilize the ligand bound conformation, but are broken in the ligand-free conformation (Figure 5). For instance in the TPP riboswitch (Figure 5B), ECs between the L5 loop and J3-2 helix (69A-37G, 69A-23C and 70A-22C, in numbering from 2gdi.pdb (Serganov et al., 2006)) reveal a set of base stacking interactions that form when TPP ligand binds, but are broken when TPP is released (Serganov et al., 2006). These functionally important interactions are completely missed by $MI_E$. Similarly, very high ranking ECs between nucleotides between P2-P4 and P1-P4 in the SAM riboswitch (Figure 5A) reveal contacts that only form conditionally on binding of the ligand *S*-adenosylmethionine, (C25-G89, U26-A88, G27-C87, G28-C86, A9-A84, A10-A84 in numbering from 4kqy.pdb (Lu et al., 2010)). Although some of these contacts are also detectable by $MI_E$, a cluster between P1-P4 are not, including base stacking between A9-A84.

### Prediction of contacts for 160 RNAs of unknown structure

Encouraged by the accuracy of contact inference for RNAs of known structure, we predicted 3D contacts for 160 RNA genes in the RFAM database that do not have a known 3D structure for any member of the family (detailed results are available online at marks.hms.harvard.edu/ev_rna/). These predictions can aid experimental structure prediction and provide 3D constraints for direct folding simulations. In the following sections we show the nucleotide resolution predictions in biological context.

### Disambiguating alternative structures of the HIV Rev Response Element RNA

To be exported from the nucleus, HIV transcripts assemble into ribonucleoprotein particles that contain multiple copies of the HIV Rev protein bound to the ~ 350 nucleotide-long RRE (Bartel et al., 1991; Iwai et al., 1992; Kjems et al., 1992; Peterson and Feigon, 1996; Rausch and Le Grice, 2015). Since the RRE and its interaction with Rev are vital steps in the HIV life cycle, they have become the subject of intense research as potential drug targets (Gallego and Varani, 2001; Luedtke and Tor, 2003; Sreedhara and Cowan, 2001), but important details of Rev-RRE structure and function remain unknown.

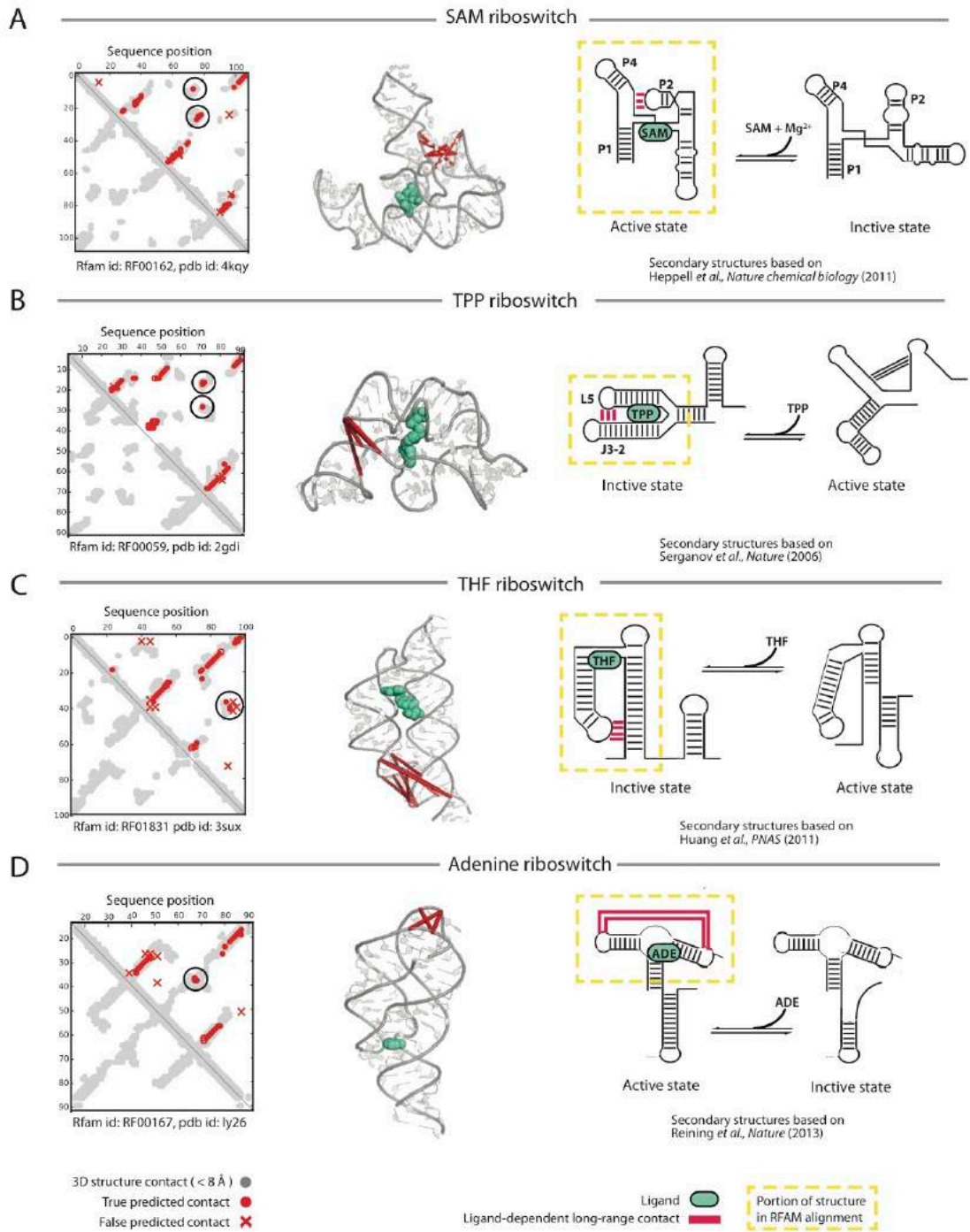

**Figure 5: Evolutionary couplings identify functional interactions in riboswitches**
3D contacts revealed by ECs are conserved across the RNA family, and may therefore be functionally important. We found that the top ranking long-range ECs in four riboswitches (A, B, C, D) are functionally critical, since they are differentially satisfied in the ligand-bound and ligand-free conformation. In each example, a contact map (left) shows the top L/2 contacts. The circled contacts – which are highlighted red on the 3D structures (middle) – are formed in the ligand-bound state, but violated in the unbound state. This is illustrated by the schematics (right), which were reproduced from prior studies.

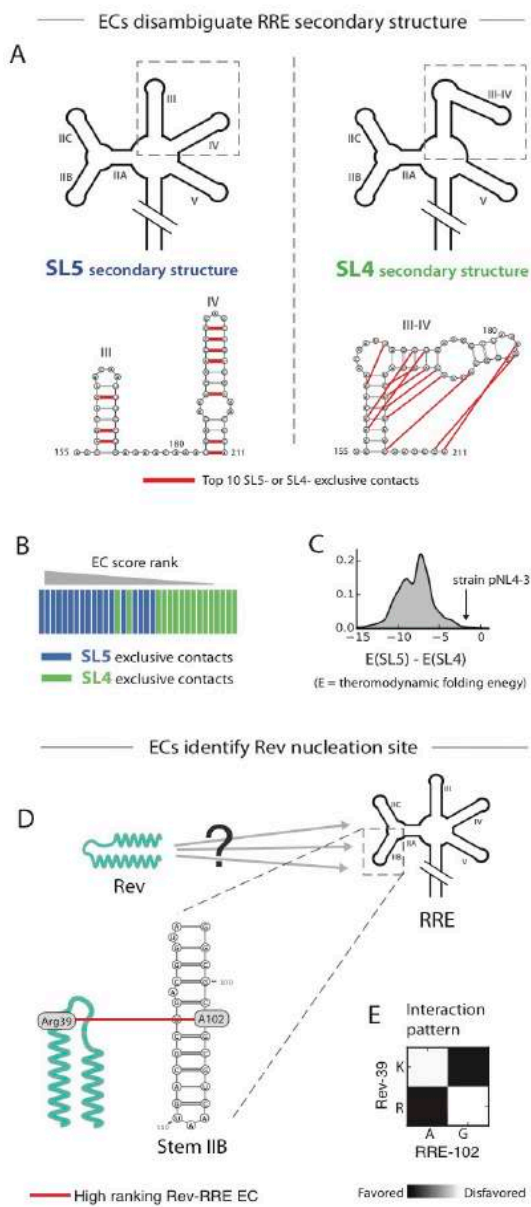

The secondary structure of RRE is one area of open investigation. *In silico* folding and *in vivo* biochemical foot-printing studies support two stable secondary structures termed SL4 and SL5, which differ in the region A155 – A211 (Figure 6A-B). SL5 has a five stem-loop structure (Bai et al., 2014; Kjems et al., 1992; Pollom et al., 2013), while recent work has shown that SL4, a four stem-loop structure, also exists both *in vitro* and *in vivo* (Charpentier et al., 1997; Fernandes et al., 2012; Legiewicz et al., 2008; Sherpa et al., 2015; Zemmel et al., 1996).

We found that ECs (Data S5) computed on the RFAM alignment of RRE (RF00036) overwhelmingly support the SL5 structure. Among 34 secondary structure contacts unique to either SL4 or SL5, the 10 highest ranked support SL5 (Figure 6A as red lines). In fact almost all of the 18 SL5-exclusive contacts outrank the 16 SL4-exclusive contacts (Figure 6B, $p < 10^{-5}$ based on t-test of EC-ranks). Although the evolutionary evidence supports SL5 we cannot rule out the existence of the SL4 secondary structure in an ensemble, but note that the sequence of the clone used by groups reporting SL4 (pNL4-3) (Adachi et al., 1986; Charpentier et al., 1997; Fernandes et al., 2012; Legiewicz et al., 2008; Sherpa et al., 2015; Zemmel et al., 1996)) thermodynamically favors SL4 more than 98.6% of the 44,046 sequences in our alignment (Figure 6C, Data S6). We obtained virtually identical results for both the thermodynamic and co-evolutionary analyses (Figure S6) using a separate multiple sequence alignment from the LANL HIV database (http://www.hiv.lanl.gov). In summary, our thermodynamic folding energy calculations and comparative sequence analysis indicate that SL5 rather than SL4 is the major RRE secondary structure, without excluding the possibility that SL4 occurs in particular evolutionary contexts, such as the pNL4-3 variant.

**Figure 6: Insight into HIV Rev Response Element (RRE) structure and function**
The HIV Rev Response Element (RRE) is an important drug target because its nuclear export function is critical to the HIV life cycle, but important details of RRE structure and function remain unknown. We first used ECs to disambiguate between two mutually exclusive RRE secondary structures reported in the literature, termed SL5 (A, left) and SL4 (A, right). We defined RRE contacts as SL4- or SL5- exclusive if they are satisfied in one secondary structure but not the other. Strikingly, 10/10 of the top-ranking exclusive contacts (red lines, bottom of A) support SL5, but not SL4. In fact almost all of the 18 SL5-exclusive contacts outrank the 16 SL4-exclusive contacts (B, $p < 10^{-5}$). Strikingly, all of the studies we found reporting the SL4 structure used the pNL4-3 variant of HIV, which is in the 98.6th percentile for favoring SL4 over SL5 (C) according to thermodynamic folding energy predictions (Data S6). Thus, SL5 is likely the dominant RRE secondary structure in most evolutionary contexts. We next scanned for evolutionarily constrained interactions between RRE RNA and Rev protein (D, top). Our top-ranking contact (D, bottom and E) linking Rev Arg39 to RRE A102 on stem IIB is consistent with biochemical work showing that stem IIB is the major Rev nucleation site. This evolutionary analysis, which reflects *in vivo* reality in diverse contexts, thus supports the *in vitro* experiments that have used only partial and synthetic RNA constructs. (See Figure S6 for identical analysis on a different RRE alignment.)

### Constrained interactions between HIV Rev Response Element RNA and Rev protein

Assembly of the Rev-RRE nuclear export complex – a key step in the HIV life-cycle – is thought to be driven by early nucleation events dependent on highly specific RNA-protein interactions. We therefore used evolutionary couplings to infer dominant points of interaction in the interface between the RRE RNA and Rev protein, an HIV encoded co-factor of nuclear export. Biochemical work shows that Rev protein oligomerizes on the RRE RNA after initial binding to stem IIB (Figure 6A) (Bai et al., 2014; Casu et al., 2013; Kjems et al., 1992). Our co-evolutionary analysis using a concatenated Rev-RRE alignment revealed a few candidate inter-molecular contacts (Data S5). Strikingly, the top scoring intermolecular evolutionary coupling (EC) pinpoints a region previously identified as the location of Rev binding to RRE in multiple experiments over the last 25 years (Bai et al., 2014; Bartel et al., 1991; Daugherty et al., 2010; Heaphy et al., 1991; Ippolito and Steitz, 2000; Malim and Cullen, 1991). This evolutionary analysis, which reflects *in vivo* reality in diverse contexts, thus supports the *in vitro* experiments that have used only partial and synthetic RNA constructs.

### Distant T box riboswitch regions coevolve via cooperative binding of a tRNA ligand

Using the inferred evolutionary couplings (ECs), we investigated the mechanism of tRNA sensing by the T box riboswitch (RF00230), a family of RNA elements that ensure homeostasis of charged tRNA by up-regulating tRNA synthetases and amino acid importers when aminoacylation of a specific tRNA is low (Green et al., 2010). T box riboswitches contain two tRNA recognition elements (Figure 7A, right): the "specifier sequence" which examines tRNA identity by base-pairing to its anti-codon sequence (Grundy et al., 2002) and the anti-terminator (T box) domain, which selectively binds the un-charged tRNA acceptor stem. These interactions stabilize the anti-terminator stem, causing a conformational change in the riboswitch that exposes a Shine-Dalgarno (SD) sequence and results in translation of the operon encoded genes (Figure 7A, right). Important details of this process have been elucidated, (Caserta et al., 2015; Grigg and Ke, 2013; Zhang and Ferre-D'Amare, 2013), but only this first stem-loop has been observed in 3D to date (Zhang and Ferre-D'Amare, 2013) it is still unclear how tRNA stabilizes the anti-terminator stem and whether the specifier sequence alone is sufficient to ensure specificity *in vivo*.

We examined the top ranking ECs for the T box riboswitch family (see Methods) to investigate the mechanism of combinatorial tRNA sensing. Remarkably, we found a cluster of six long-range ECs connecting the specifier sequence to the T box region (pink arc in Figure 7A) that are likely mediated by co-variation with the intervening tRNA. In particular, variation in the nucleotides of the specifier sequence may have led to changes in tRNA specificity, which in turn affected the evolution of nucleotides in the T box region, producing the observed co-variation. If true, this hypothesis would imply that the T box region collaborates with the specifier sequence to establish tRNA specificity, which is consistent with mutational studies that showed interdependence between the Bacillus subtilis tyrS T box and tRNA(Tyr)-acceptor sequences (Grundy et al., 1994).

We next looked at which specific nucleotides from the T box region participate in the six long-range ECs with the specifier sequence. Strikingly, two of the six long-range ECs involve a T box nucleotide (U191, RFAM reduced numbering) that base pairs to the tRNA discriminator position (Crothers et al., 1972), which has been known for decades to co-vary with the tRNA anti-codon (Klingler and Brutlag, 1993) and is thought to influence the tertiary structure of the acceptor stem (Lee et al., 1993). The other four long-range ECs involve bases G186 and C210 that are paired in the anti-terminator stem, raising the possibility that the tRNA acceptor stabilizes the anti-terminator stem by forming extensive tertiary contacts with bases in and near the T box region. Using an adaptation of our model that previously demonstrated accurate inference of the effects of single-site mutations (Hopf et al., 2015) (see Methods), we predict that in the Enterococcus durans T box, for example, the mutations G186U/C210A and U191A would be most likely to disrupt these interactions.

## Proteins displace RNA tertiary contacts in the evolution of RNase P

We next used evolutionary couplings (ECs) to address a longstanding hypothesis on the evolution of RNase P, an ancient ribozyme that has become a model system for investigating evolutionary plasticity of molecular structure (Krasilnikov et al., 2004). Whereas the bacterial RNase P (RF00010) can function as pure RNA, the widely diverged archeal (RF00373) and eukaryotic (RF00009) RNase P require their protein subunits (Gopalan, 2007). It is generally believed that during the evolution of RNase P in archaea and eukaryotes, the newly added protein subunits began to replace RNA to stabilize the overall tertiary architecture and establish the active site (Evans et al., 2006). However, a lack of knowledge of RNA-RNA intra-molecular interactions in the archeal and eukaryotic RNase P – which have no known structure – has made it difficult to critically evaluate this hypothesis about their evolution.

Given our high prediction accuracy for RNA-RNA contacts (94% precision for detecting base pairs within 12 Å) in the bacterial RNase P (Figure 7B, left), whose structure is known, we investigated whether our predicted contacts for the archeal and eukaryotic RNase P support the hypothesis that proteins have replaced RNA to establish key tertiary interactions (see Methods for selection of contacts). Consistent with this hypothesis, we found several coevolving pairs of nucleotides from the bacterial RNase P that do not correspond to any high ranking ECs in the archeal or eukaryotic lineage, but emanate from helices that are now the sites of protein binding in eukaryotes (Figure 7B, right). For example, contacts (3A-95G, 95G-341U, in the numbering of 3q1q.pdb (Reiter et al., 2010)) connecting P1 to P9 in bacteria are not predicted for the eukaryotic RNase P, where P9 now binds the protein Pop4 (Khanova et al., 2012). Similarly, we predicted contacts (43C-299C, 44C-299C, 23C-299C) from P3 to P18 in bacteria, but not in eukaryotes, where P3 now binds the protein Pop7 (Khanova et al., 2012).

Alongside predictions that support an increased role for proteins in establishing the tertiary structure of eukaryotic RNase P, we also find high ranking ECs that do not support this role, but instead suggest that RNA-RNA tertiary contacts have emerged *de novo* in the eukaryotic lineage to compensate for the loss of key secondary structural elements. These contacts, including G5-A86 (RFAM reduced numbering) from P1 to 3' of P8 and C4-G234 from P1 to conserved region IV (CR IV), significantly coevolve in eukaryotes, but correspond to no high ranking ECs in bacteria. Notably, even though the bacterial homologs of these base pairs do not significantly coevolve, they are nevertheless close in 3D space (< 2 nm).

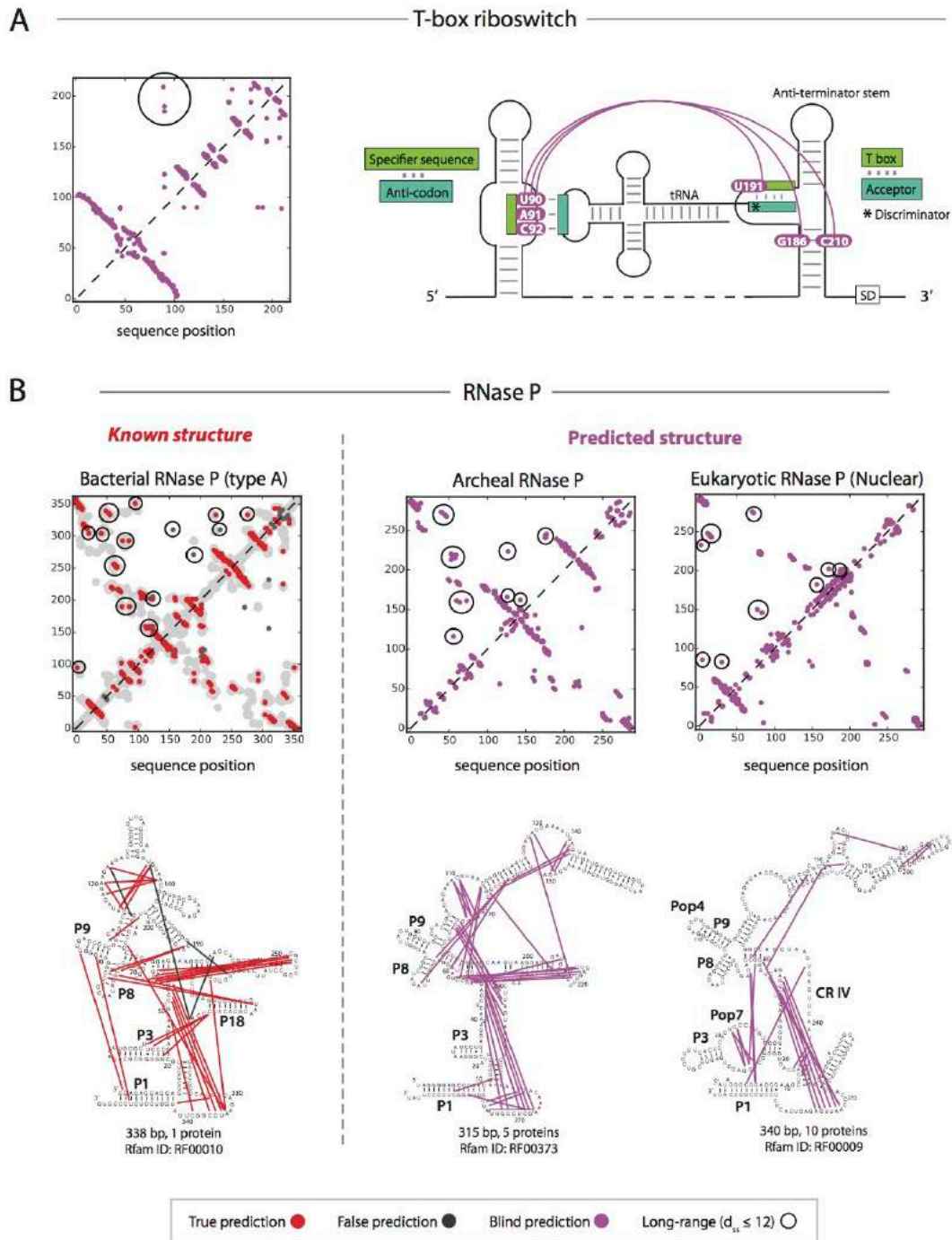

**Figure 7: Examples of evolutionary couplings elucidating structure and function**
We provide ECs for 160 RFAM families without a known structure, and investigate them in detail for the T box riboswitch (A) and RNAs P (B). In the T box riboswitch, a high-ranking cluster of long-range ECs (A, circled dots in the contact map and pink arcs on the secondary structure) connects the specifier sequence to the T box region. The binding of both of these RNA elements to an un-charged tRNA induces translation of downstream genes by exposing a Shine-Delgarno (SD) sequence. The long-range ECs are probably mediated by co-variation with the intervening tRNA. Supporting this hypothesis, two of the six long-range ECs involve U190, a nucleotide in the T box region that base pairs to the discriminator position in the tRNA (A, marked by the *), which is known to co-vary with the tRNA anti-codon. (B) We also used ECs for the bacterial, archeal and eukaryotic RNase P to address a hypothesis in the evolution of RNase P, a ribozyme found in all three domains of life.

## Discussion

We show that evolutionary couplings (ECs) derived from sequence co-variation using a global maximum entropy model can predict 3D contacts in RNA, including long-range tertiary contacts and non Watson-Crick base pairs, with good accuracy. These contacts allow blinded 3D structure prediction using only sequence information with accuracy matching the state of the art with biochemical probing data. ECs not only provide a window into RNA structure, but also can reveal functionally important interactions such as tertiary contacts that distinguish the ligand-bound from the ligand-free conformation of four riboswitches or the potentially complex-nucleating interaction between the RRE RNA and the Rev protein of HIV virus. Since they are computed from broad alignments of homologous sequences, EC-inferred contacts are likely stable in diverse evolutionary contexts and conserved across evolution. Thus, in contrast to unbiased experimental approaches, 3D information from ECs has the unique potential to highlight functionally important interactions in structured RNA.

Building on the success of ECs at detecting contacts in RNA and protein separately, we provide proof of principle for the detection of intermolecular contacts in RNA-protein complexes from evolutionary sequence information alone. High-throughput elucidation of RNA-protein interactions from sequence alone would be a major breakthrough, but requires the development of improved bioinformatics approaches for curating and phasing RNA-protein alignments.

We identify several growth areas for our approach. One shortcoming is that we always report the top L/2 contacts (where L is the length of the RNA), whereas flexibly choosing the number of contacts would better suit the natural diversity of 3D contact abundance in different RNAs. Further improvements can be made in folding pipeline, which is currently not optimized for large RNAs (> 120nt), despite the fact the evolutionary couplings yield highly accurate contacts for RNAs as large as the ribosome. Potential changes include higher sampling density, use of negative restraints, multi-stage annealing with continual addition of contacts, better ranking of decoy structures, and better filtering of false positive contacts.

Another area for future research is to build on the outstanding work of the RFAM database (Nawrocki et al., 2015) by extending sequence coverage, and including more transcripts that potentially have 3D structure. Though detection of accurate ECs for RNA requires far less sequence coverage than proteins due to its smaller alphabet size and lower average number of contacts per residue, sequence abundance is still the rate-limiting component of our approach. Fortunately, the ongoing explosion of available sequence data means that the outlook elucidating of functional interactions in mRNAs, lncRNAs, viral genomes – and their protein-binding partners – is promising.

# Experimental Procedures

## Calculating evolutionary couplings (ECs)

To calculate ECs, we fit a global probability model to each alignment using pseudo-maximum likelihood, as described below. Our full pipeline as single script is available here: https://github.com/debbiemarkslab/plmc.

1. Sample re-weighting (and definition of $M_{eff}$)

To prevent phylogenetic correlations between sequences from biasing our analysis, we down-weight each sequence $\sigma$ by the number $N(\sigma)$ of neighbors it has in sequence space. We then define the "effective size" of the alignment as the sum of weights:

$$M_{eff}(A) = \sum_{\sigma \in A} \frac{1}{N(\sigma)}$$

2. Model fitting

We fit the following probability model using a pseudo-maximum likelihood approximation (Besag, 1975) and $L_2$ regularization.

$$P(\boldsymbol{\sigma}) = \frac{1}{Z} \exp\left( \sum_{i=1}^{L} h_i(\sigma_i) + \sum_{i=1}^{L-1} \sum_{j=i+1}^{L} J_{ij}(\sigma_i, \sigma_j) \right)$$

The $h_i$ terms represent single-site conservation and the $J_{ij}$ represent co-variation.

3. Post-processing

Once the parameters $\boldsymbol{h}$ and $\boldsymbol{J}$ have been fit to data, we take the Frobenius norm $FN(i,j)$ of the $J_{ij}$ couplings. In theory, the resulting $FN$-scores represent the strength of co-variation at each position. However, under-sampling effects lead to a characteristic distortion of these scores, which we correct using an average product correct (APC) (Dunn et al., 2008). APC-corrected $FN$-scores are reported as the final ECs.

## 3D structure prediction

We predicted all-atom sturctures for five RNA families, representing the subset that (i) have a known structure (ii) have length between 70-120nt (iii) have at least one *highly-long-range* contact, defined as a contact with $d_{ss} \geq L/4$, where L is the length of the RNA. We performed structure prediction with Nucleic Acid Simulation Tool (NAST) (Jonikas et al., 2009), a coarse-grained modeler that uses a combination secondary structure and tertiary contacts as inputs (Figure S7). For each RNA family, we created 600-1000 decoy coarse-grained models and then clustered the 20% of decoys with the lowest energy-per- contact using k-means with k = 4. The final 4 candidate models were then given all-atom structures and refined in XPLOR (Schwieters et al., 2003). All 4 models for each family are shown in Figure S3, and the ones with closest match to the experimental structure are shown in Figure 3.

## RNA-protein structure prediction

To compute RNA-protein ECs, we used the same approach as for RNA (described above) but now with a full alphabet including all amino acids. No other changes were made to the model. To determine whether EC-derived RNA-protein contacts improve 3D structure prediction of RNA-protein complexes, we used these contacts as restraints for rigid body docking in HADDOCK (Dominguez et al., 2003). For docking controls, we applied center of mass restraints only.

## Supplementary Experimental Procedures

### Selection of RFAM families

To test whether ECs could predict tertiary structure contacts, we used RNA multiple sequence alignments from the RFAM 11.0 database (Burge et al., 2013), removing columns with > 50% gaps. Mappings from the original RFAM coordinates to "RFAM reduced coordinates" – in which gaps have been removed – are recorded in Data. We restricted to families where the effective number of sequences ($M_{eff}$, see below) was greater than 0.5L, where L is the number of columns in the alignment, yielding 182 families (see https://marks.hms.harvard.edu/ev_rna/). Of these, 22 aligned to a known structure in the PDB (Berman et al., 2000). Ranked lists of ECs are contained in Data S1

### Computing evolutionary couplings

*Summary*
To identify co-evolving nucleotides in RNA alignments, we fit a global statistical model to the sequences that is parameterized by single-site bias and pairwise coupling terms. In contrast to models of co-variation that consider pairs in isolation, such as mutual information (MI), this global statistical model can de-convolve transitive, chained co-variation into a typically smaller, more concentrated set of underlying couplings. In the following sections we (i) present the probability model; (ii) outline an approximate penalized Maximum Likelihood approach for fitting the model; and then describe three additional features that improve prediction accuracy, including (iii) regularization, (iv) sample reweighting; and (v) average product correction. Code is available at https://github.com/debbiemarkslab/plmc.

*(i) Description maximum entropy probability model*
We model the probability of a sequence $\boldsymbol{\sigma} = (\sigma_1, \ldots, \sigma_L)$ of length $L$ as

$$P(\boldsymbol{\sigma}) = \frac{1}{Z} \exp\left( \sum_{i=1}^{L} h_i(\sigma_i) + \sum_{i=1}^{L-1} \sum_{j=i+1}^{L} J_{ij}(\sigma_i, \sigma_j) \right).$$

The external fields $h_i$ represent single-site conservation and the pair couplings $J_{ij}$ represent co-variation. For example, the term $J_{ij}(\sigma_i, \sigma_j)$ represents the statistical energy contributed by nucleotide $\sigma_i$ in position $i$ interacting with nucleotide $\sigma_j$ in position $j$. Thus, if there are $L$ nucleotides total, with each taking 5 possible states (A,C,G,U and gap), then $\boldsymbol{J}$ can be thought of as a $L \times L \times 5 \times 5$ matrix, where each 5×5 slice describes the pattern of co-variation between a given pair of positions.

The *partition function Z* ensures that *P* is properly normalized, and is given by

$$Z = \sum_{\sigma} \exp\left( \sum_{i=1}^{L} h_i(\sigma_i) + \sum_{i=1}^{L-1} \sum_{j=i+1}^{L} J_{ij}(\sigma_i, \sigma_j) \right)$$

Once the parameters $\boldsymbol{h}$ and $\boldsymbol{J}$ have been fit to data, we use the Frobenius norm $FN(i,j)$ of the $J_{ij}$ couplings to assess the strength of coupling between position $i$ and $j$, as follows.

$$FN(i,j) = \| J_{ij} \|_2 = \sqrt{\sum_k \sum_l J'_{ij}(k,l)^2}$$

where $J_{ij}'$ is a centered version of $J_{ij}$ with row and column means set to 0. The *FN* scores are used to generate evolutionary couplings (ECs), as described in part (v).

*(ii) Model fitting by pseudo-maximum likelihood (PLM)*
A standard, consistent method for inferring the parameters of probability models is maximum likelihood, where the parameters are chosen to maximize the probability of the observed data under the model. Direct maximum likelihood is ill suited to the model described above, since it requires computing $Z$ directly, which is intractable. Instead of maximizing the likelihood, we instead maximize a surrogate function, the pseudolikelihood (Besag, 1975) which approximates the full likelihood for each sequence $\boldsymbol{\sigma} = (\sigma_1, \ldots, \sigma_L)$ by a product of conditional likelihoods for each site $i$:

$$P(\sigma_1, \ldots, \sigma_L \mid \boldsymbol{h}, \boldsymbol{J}) \approx \prod_{i=1}^{L} P(\sigma_i \mid \boldsymbol{\sigma} \setminus \sigma_i, \boldsymbol{h}, \boldsymbol{J})$$

By computing a likelihood for each site $i$ while conditioning on the remainder of the sequence ($\boldsymbol{\sigma} \setminus \sigma_i$), the global partition function $Z$ is replaced by a number of local partition functions, so that all terms in the approximate likelihood function (shown below) are tractable.

$$P(\sigma_i | \boldsymbol{\sigma} \setminus \sigma_i, \boldsymbol{h}, \boldsymbol{J}) = \frac{\exp(h_i(\sigma_i) + \sum_{j \neq i} J_{ij}(\sigma_i, \sigma_j))}{\sum_a \exp(h_i(a) + \sum_{j \neq i} J_{ij}(a, \sigma_j))}$$

This pseudolikelihood approach has previously been applied (Ekeberg et al., 2013; Hopf et al., 2015; Hopf et al., 2014; Kamisetty et al., 2013; Ovchinnikov et al., 2014) to estimate residue couplings in protein sequence families. We optimize this approximate likelihood function (with some modifications outlined in (iii) and (iv)) using a quasi-Newton method (L-BFGS).

### *(iii) Regularization*

The number of parameters to estimate in $\boldsymbol{J}$ is $\frac{L(L-1)}{2} q^2$, where $L$ is the length of the sequence and $q$ is the number of states (i.e. ~$10^6$ parameters for an RNA of length 200). Since this vastly exceeds the effective number of sequences in a typical alignment, parameters must be strongly regularized to limit over-fitting. To that end, we use $L_2$-regularization of the fields $\boldsymbol{h}$ and couplings $\boldsymbol{J}$ with strength $\lambda_h$ and $\lambda_J$ respectively:

$$\mathcal{R}(\boldsymbol{h}, \boldsymbol{J}) = \lambda_h \sum_{i=1}^{L} \| h_i \|_2^2 + \lambda_J \sum_{i=1}^{L-1} \sum_{j=i+1}^{L} \| J_{ij} \|_2^2$$

Optimizing this augmented objective involves a tradeoff between fitting the data (by increasing the pseudo-likelihood) and maintaining small parameter values (by decreasing the regularization term $\mathcal{R}$). If $\mathcal{L}(\boldsymbol{h}, \boldsymbol{J})$ denotes the pseudo-likelihood of parameters $\boldsymbol{h}$ and $\boldsymbol{J}$ under the data, then we can write the optimization problem as follows.

$$\boldsymbol{h}, \boldsymbol{J} = \arg \min_{hJ} \left( -\log \mathcal{L}(\boldsymbol{h}, \boldsymbol{J}) + \mathcal{R}(\boldsymbol{h}, \boldsymbol{J}) \right)$$

In this work, we set the regularization strength as $\lambda_h = 0.01$ and $\lambda_h = 20.0$ for all computations of ECs (i.e. both for RNA alone and RNA-protein together).

### *(iv) Sample reweighting*

Our maximum entropy approach models the sequences in an alignment as independent draws from an underlying distribution. However, this assumption does not hold in reality, since sequences are usually related by phylogeny. To account for this, we reweight sequences in inverse proportion to the size of their sequence neighborhood. Formally, a sequence $\sigma$ of length $L$ in alignment $A$, is given the weight

$$w(\sigma) = \frac{1}{m(\sigma)} \quad \text{where} \quad m(\sigma) = |\{\sigma' \in A \mid \text{hamming}(\sigma, \sigma') * L^{-1} < \theta\}|$$

$\theta$ is a user-defined parameter determining neighborhood size. In this study, we used $\theta = 0.2$ for all RFAM alignments and $\theta = 0.1$ for RNA-protein alignments, except for alignments involving the HIV genes Rev and RRE, for which used $\theta = 0.033$.

The sum of weights $w(\sigma)$ over all sequences in the alignment represents the total *effective* number of sequences ($M_{\text{eff}}$). In other words,

$$M_{eff}(A) = \sum_{\sigma \in A} w(\sigma)$$

*(v) Average product correction*

The *FN* scores defined in part (i) cannot be directly used for inferring structure contacts, since they are contaminated by bias due to phylogeny and undersampling. Fortunately, these artifacts are concentrated in the top eigenvector of the *FN* matrix, and can therefore be removed using an average product correction (APC), which reconstitutes the *FN* matrix from its eigenvectors while setting the top eigenvalue to 0. The resulting APC-corrected *FN* matrix contains the evolutionary coupling (EC) scores referenced throughout this paper.

In practice, we apply the APC by subtracting normalized row and column averages from each position, as follows.

$$EC(i,j) = FN(i,j) - \frac{\left(\sum_{i' \neq i} FN(i',j)\right)\left(\sum_{j' \neq j} FN(i,j')\right)}{\sum_{i'} \sum_{j' \neq i} FN(i',j')}$$

## Prediction of mutational effects

Our predictions of mutations likely to disrupt key interactions – presented for the T box riboswitch and RNase P – are based on the inferred parameters **h** and **J** of the global probability model (see above) and follow the methodology presented in (Hopf et al., 2015). Briefly, consider a sequence $\sigma$ and mutation *m*. Let $\sigma'$ be the sequence that results from applying *m* to $\sigma$. The effect of mutation *m* is calculated as

$$\text{Effect}(m) = E(\sigma') - E(\sigma) \quad \text{where} \quad E(\sigma) = \sum_{i=1}^{L} h_i(\sigma_i) + \sum_{i=1}^{L-1} \sum_{j=i+1}^{L} J_{ij}(\sigma_i, \sigma_j)$$

## Computing MI

To investigate how ECs compare to previous measures of co-evolution, we computed two versions of mutual information (MI). First we computed the raw MI ($MI_R$) as shown below, where $f_i(A) = P(S_i = A)$ and $f_{ij}(A,B) = P(S_i = A, S_j = B)$ for a sequence *S* in the alignment.

$$MI_R(i,j) = \sum_{A,B} \frac{f_{ij}(A,B)}{f_i(A)f_j(B)}$$

EC scores differ from $MI_R$ in three ways: (1) They rely on a global maximum entropy model; (2) They down-weight sequences with a greater phylogenetic representation in the alignment; (3) They include an APC correction. Since feature (1) is the focus of this study, we also computed an enhanced MI score ($MI_E$), which incorporates features (2) and (3), as has been done in previous work on RNA co-evolution (Dunn et al., 2008).

## Annotating interactions

For each alignment, we investigated the top L/2 contacts with chain-distance > 4. We first classified contacts as true-positives if the minimum-atom-distance from the crystal structure was < 8 Å. These were classified according secondary structure distance ($d_{ss}$) and biochemical interaction type, with long-range contacts defined as those satisfying $d_{ss}$ > 4. The $d_{ss}$ for a pair of bases is the length of the shortest path between them in a graph where nodes are bases and edges are either secondary-structure contacts or instances of adjacency on the chain. To compute $d_{ss}$, we used the consensus secondary-structure provided by RFAM, which is inferred using a profile stochastic context-free grammar (Nawrocki and Eddy, 2013). To classify contacts by their biochemical interaction type, we used crystal structure annotations from FR3D (Petrov et al., 2011; Sarver et al., 2008) which were downloaded from RNA3DHub (http://rna.bgsu.edu/rna3dhub/). Ranked lists of ECs for each of the 22 RFAM families with a matching PDB structure are provided in Data S1.

## Computing 3D structures from evolutionary couplings

We performed blinded structure prediction for all RNA families that (i) Have a known structure (ii) Have length between 70-120nt (iii) Have at least one *highly-long-range* contact, defined as a contact with $d_{ss} \geq L/4$, where L is the length of the RNA. We performed structure prediction with Nucleic Acid Simulation Tool (NAST) (Jonikas et al., 2009), a coarse-grained modeler that uses a combination secondary structure and tertiary contacts as inputs, followed by refinement in XPLOR (Schwieters et al., 2003). We describe the folding pipeline in detail below.

1) For each RNA family, we generated 200 random unfolded structures that satisfied the secondary structure constraints (Figure S7A).
2) Next, we performed molecular dynamics using tertiary structure restraints to generate candidate models (Figure S7B) with a restraint energy of 40. To obtain these tertiary structure restraints, we used the N long-range contacts with the top EC scores, where N varied between 20 and the length L of the RNA in intervals of 20. Thus, for a typical RNA family, we used around 4 different restraint sets, where the first set had the least contacts and the fourth had the most.
3) Since many restraint sets contained false-positives (i.e. restraints that are not satisfied in the true 3D structure), we used an iterative pruning procedure to remove contacts that were not consistent with the rest of the set. To that end, we performed molecular dynamics using weak constraints to iteratively remove restraints that were consistently violated by the resulting structures (Figure S7C), removing at most 15% of the contacts in any one round. Contacts were defined to be violated when the average distance between the corresponding bases was > 15 Å. At the end of this process, each restraint set from part (2) had been replaced by a subset, where all the restraints in the subset were consistent with each other.
4) At the end of steps (1-3), we obtained 200 decoy models for each restraint set, meaning 600-1000 decoys for each RNA family (longer RNAs had more restraint sets and therefore more decoys). We then assigned to each decoy an energy-per-contact, defined as $E/N$ where E is NAST energy of the decoy and N is the number of contacts in its restraint set. For each RNA family, we then clustered the 20% of decoys with the lowest energy-per-contact using k-means with k = 4. RMSDs were calculated using Biopython, (Figure S7D). See Figure S4 for plots of energy vs. RMSD.

5) From each cluster, we chose a lowest energy representative and then created an all-atom structure using the NAST C2A pipeline (Figure S7E).
6) Finally, we refined the all-atom models by simulated annealing with XPLOR (Figure S7F). Thus, at the end of this pipeline, we produce four candidate predicted structures for each RNA family.

To analyze the predicted structures for each RNA family, we calculated the all-atom RMSD from the true structure using Pymol and used Molprobity (Davis et al., 2004) to analyze the produce scores quantifying the structures' intrinsic quality, i.e. how much they reproduce the geometry of 'typical' RNA structures. The predicted structure with lowest RMSD to the crystal structure is shown in Figure 3. In addition, we ran the above pipeline with no tertiary restraints as a control (see Figure 3B). RMSD values and Molprobity scores are available in Data S2.

## RNA-protein 3D structure prediction

### Selection of RNA-protein complexes

To compute evolutionary couplings between a pair of interacting genes, one must accurately phase the corresponding alignments (i.e. match sequences from one alignment with sequences from the other). Previous work detecting evolutionary couplings in protein-protein complexes (Hopf et al., 2014) has benefitted from co-operonic interaction partners, which can be accurately phased using genomic distance. Since RNA-protein complexes are typically not co-operonic, we limited our analysis to universally conserved, highly specific interactions between RNAs and proteins with no close paralogs. After excluding complexes with low sequence diversity ($M_{eff} / L > 0.25$) and those that share an interface with a third interaction partner, we arrived at a final validation set of 21 RNA-protein complexes (see Data S3 for ranked EC lists).

### Obtaining protein alignments

RNA alignments for all RNA-protein complexes were taken from RFAM (Burge et al., 2013). Protein alignments were taken from PFAM (Finn et al., 2014) where it covers the full protein e.g. for RNaseP protein, and otherwise created by searching the UniProt (UniProt, 2015) database (release 2015_02) using 5 iterations of jackhammer (Finn et al., 2011), using an e-value, including columns with less than 50% gaps and removing sequences that had <50% length coverage relative to our query sequence. All resulting protein alignments are provided in Data S7.

### Concatenation

Detecting coevolution in an RNA-protein complex requires phasing or "concatenating" the sequences in the alignments of the RNA and protein respectively. We used the NCBI taxonomy ID to concatenate RNA and protein sequences, as follows. First, we identified the set of NCBI taxonomy IDs with at least one representative in both the RNA and protein alignments. Next, for IDs with more than one RNA or protein representative, we computed the average hamming distance between representatives and discarded taxonomy IDs for which the average hamming distance exceeded 1%. Thus, the remaining taxonomy IDs each had one or more highly similar RNA representatives, and one or more highly similar protein representatives. For each of these remaining IDs, we randomly chose an RNA representative and a protein representative, which together formed a line in the final alignment.

*Calculating ECs*

To compute RNA-protein ECs, we used the same approach as for RNA (described earlier in Methods) but now with a full alphabet including all amino acids. No other changes were made to the model.

*Rigid body docking*

To determine whether EC-derived RNA-protein contacts improve 3D structure prediction of RNA-protein complexes, we used these contacts as restraints for rigid body docking in HADDOCK (Dominguez et al., 2003). Specifically, we docked the 6 (out of 21) RNA-protein complexes that had least 75% true positives for the top 4 contacts. We inputted these top 4 contacts as unambiguous distance restraints ($5\pm2$ Å) into HADDOCK, and otherwise used default parameters. For docking controls, we applied center of mass restraints only. By default, HADDOCK clusters docked decoys and ranks the representatives of each cluster. We extracted the highest-ranking representative from each cluster for downstream analysis. All input models, restraint files, and output cluster representatives for cases and controls are provided in Data S7.

*Computing i-RMSD*

Interface-RMSD (i-RMSD) is a commonly used metric for assessing prediction accuracy of molecular complexes. I-RMSD is equivalent to standard RMSD, but taken over the interface between subunits, defined as the subset of atoms that lie within 10 Å (where distance is computed with respect to the experimental structure) of the subunit that they are not directly part of. We used Biopython to identify interface atoms and computed all-atom RMSDs in pymol (with no atom rejection).

**Evolutionary couplings for HIV Rev Response Element (RRE)**

We computed evolutionary couplings RRE using the RFAM alignment (RF00036) and also a custom alignment (referred to here as LANL) using sequences from Los Alamos HIV sequence database http://www.hiv.lanl.gov. To form the custom alignment, we downloaded *env* nucleotide sequences (aligned to the reference HXB2 genome). We then realigned these sequences with cmalign (Nawrocki and Eddy, 2013) using the cm profile from the reference RRE RFAM entry. *Rev* sequences (also downloaded from http://www.hiv.lanl.gov) were realigned by iterative alignment with hmmalign (Eddy, 1998) using the --hand option. To infer Rev-RRE inter-molecular contacts, we phased the LANL RRE and Rev alignments by matching sequences with same Genbank ID.

The RNA secondary structure of the SL4 RRE conformation originally proposed in (Mann et al., 1994) and the SL5 conformation were taken from (Sherpa et al., 2015) using the pNL4-3 (Genbank AF324493) genome. Energy calculations were performed using RNAeval (Lorenz et al., 2011; Mathews et al., 2004; Turner and Mathews, 2010) with the default settings from the online server http://rna.tbi.univie.ac.at/cgi-bin/RNAeval.cgi.

### Evolutionary couplings for T box riboswitch and RNase P

Contacts for the T box riboswitch (RF00230) and RNase P family members (RF00010, RF00009 and RF00373) – presented in Figure 7 – were drawn from the top L ECs with a chain distance > 4. Given the large size of these sequences, we defined contacts as long-range when they satisfied $d_{ss} \geq 12$. For the archeal RNase P (RF00373), these criteria produced a set of relatively low ranking contacts that appeared in isolation at apparently random positions in the contact map, a hallmark of false-positives. Therefore, for RF00373, we removed contacts with rank > L/2 unless they were reinforced by at least one other contact, where two contacts are considered mutually reinforcing if their endpoints are both within 1 bp of each other. Contacts mentioned in the text are given in terms of their RFAM-reduced coordinates unless stated otherwise. The six long-range contacts referred to in the discussion of the T box riboswitch are (91,191), (92,192), (91,186), (90,210), (91,210) and (90,186) in RFAM reduced numbering.

## Supplementary Figures

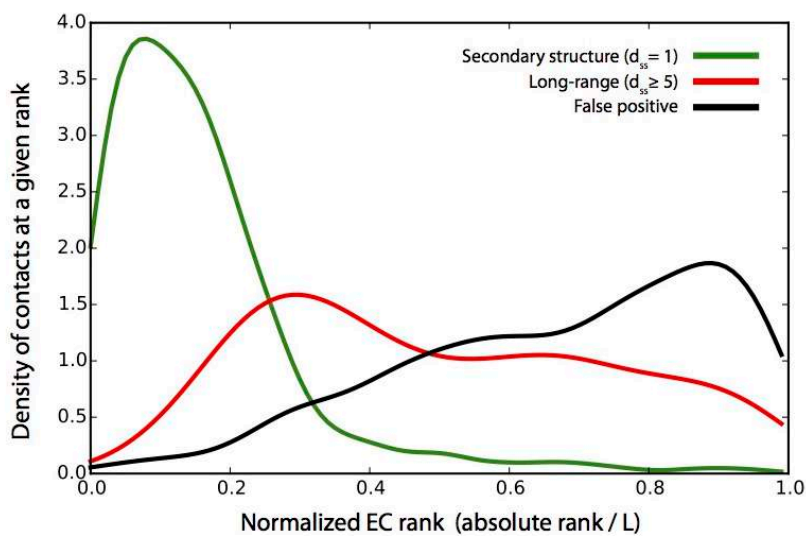

**Figure S1: EC strength for short- and long-range contacts (related to Figure 2)**
Each curve represents a histogram of normalized EC ranks for contacts of the given category. Normalized rank is R/L where R is the absolute rank and L is the number of columns in the alignment. Overall, secondary structure contacts have very high EC rank, whereas long-range contacts have a broader distribution of ranks, with some long-range contacts having higher EC score than some secondary structure contacts.

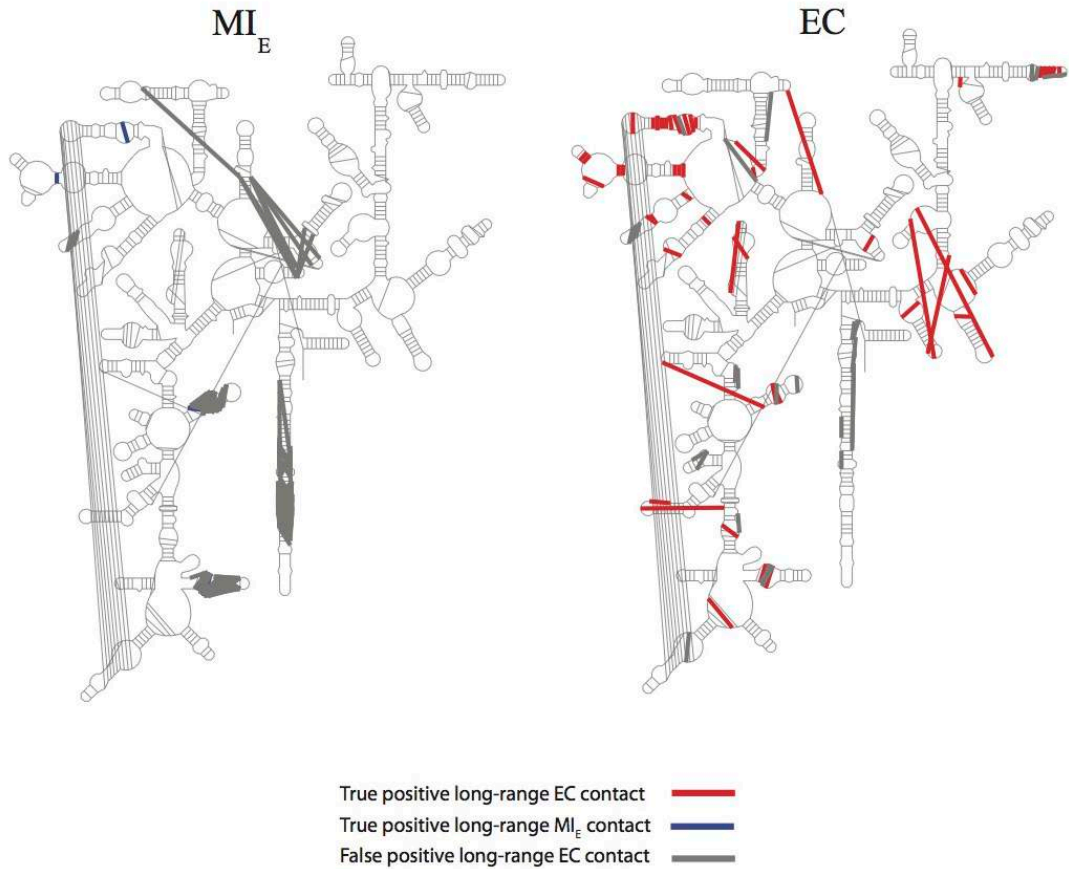

**Figure S2: Long-range contacts in the eukaryotic ribosome (related to Figure 2)**
Long-range contacts from the top L/2 predictions from EC and $MI_E$ are displayed on the secondary structure of the small subunit of the eukaryotic ribosome (RF01960). True positive contacts are colored (ECs in red, $MI_E$ in blue) and false-positives are shown in gray. Note the MI predicts dramatically fewer true-positive long-range contacts than ECs, and also 2.8-fold more false-positive long-range contacts.

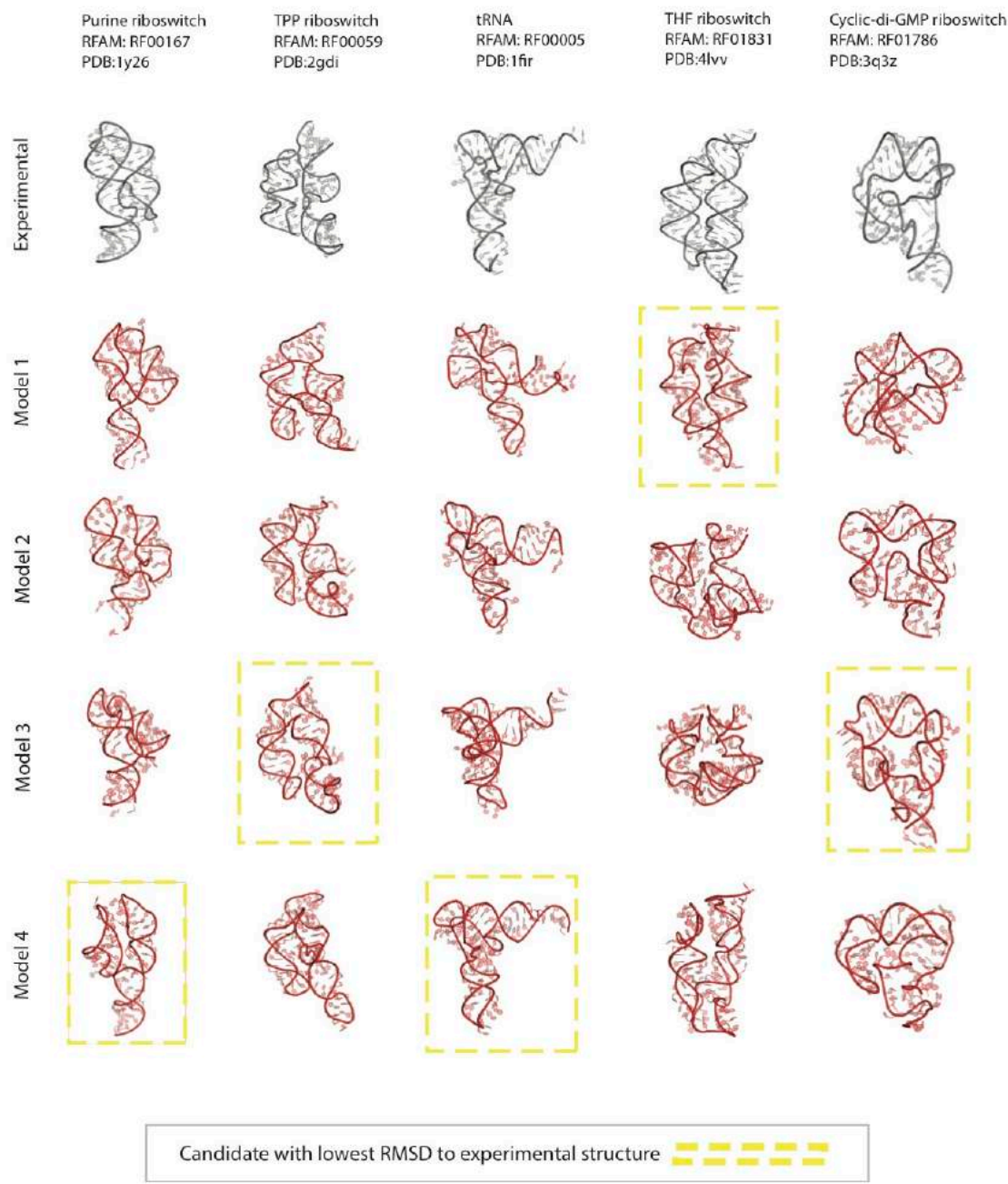

**Figure S3: Full results from RNA 3D structure prediction (related to Figure 3)**
We generated 3D structure predictions using EC-derived long-range contacts for five RNA families with a known structure. For each family, we generated four candidate predictions, each representing a cluster of decoys. All four candidates (in red) are shown in comparison to the experimental structure (gray), with the most accurate candidate highlighted in yellow.

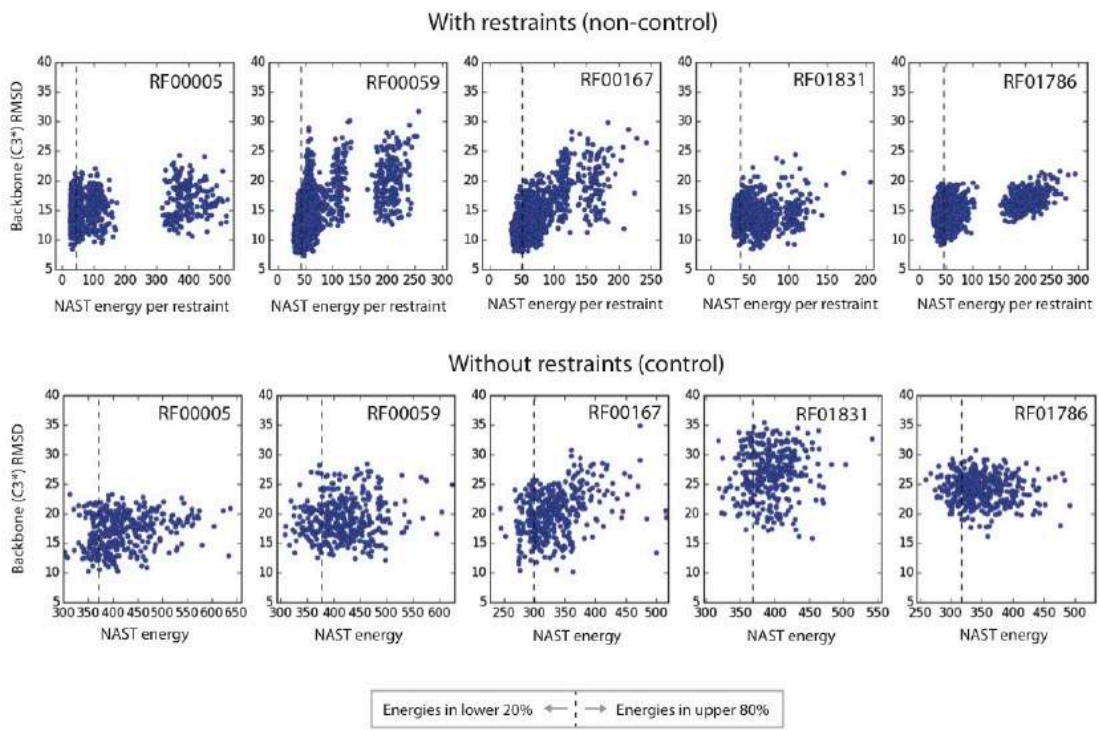

**Figure S4: RMSD vs. energy for RNA 3D structure decoys (related to Figure 3)**
For each of five RFAM families with known structure, we generated decoy models in NAST using EC-derived distance restraints (cases) and no restraints (controls). Each decoy is a point in one of the above scatterplots.

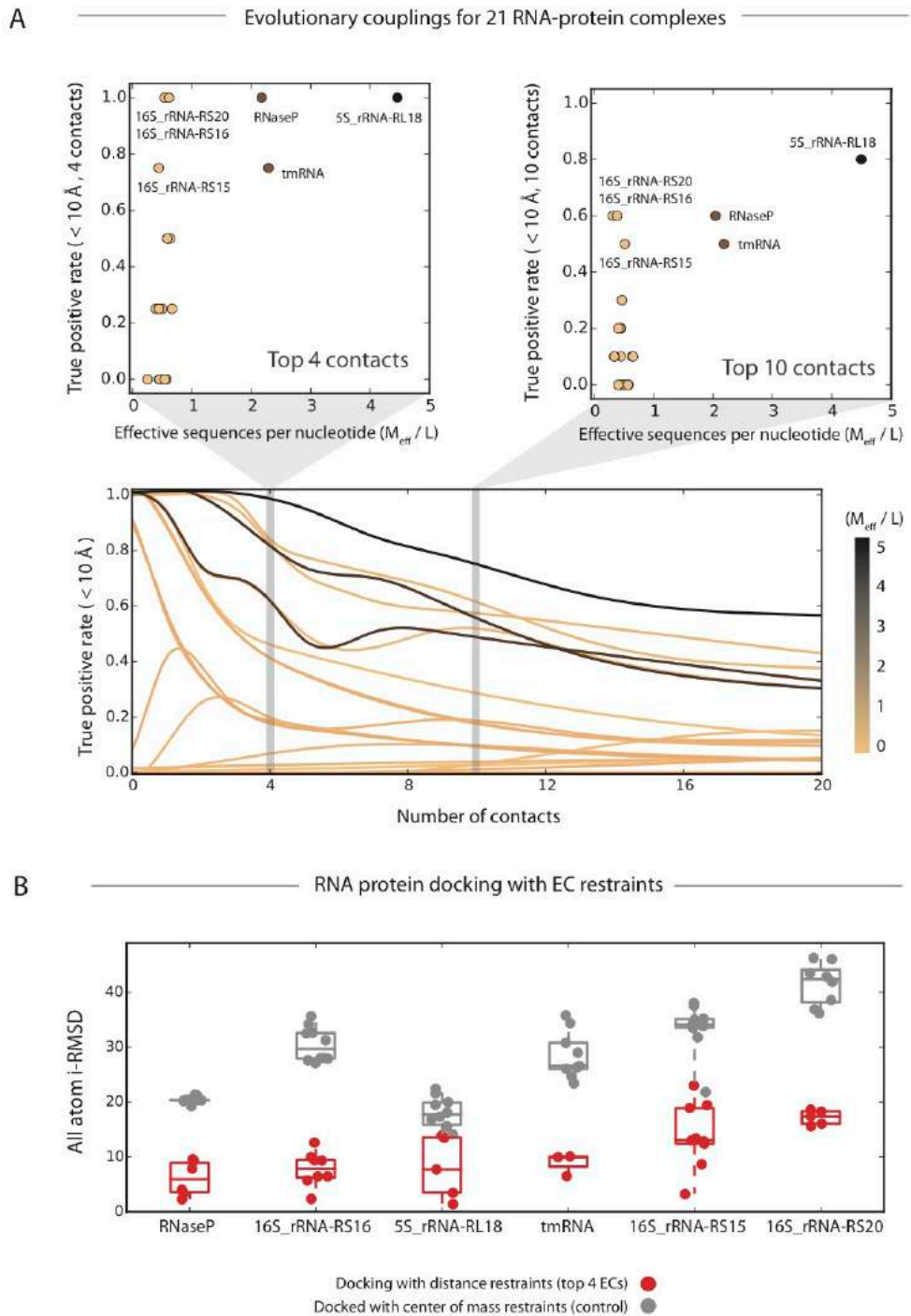

**Figure S5: ECs and docking of RNA-protein complexes (related to Figure 4)**
We computed evolutionary couplings for 21 RNA-protein complexes with known structure. (A, bottom) True positives rates for increasing numbers of contacts are overlaid, colored by the number of effective sequences per nucleotide/residue ($M_{eff}/L$). (A, top) Scatter plots of TP rate verses $M_{eff}/L$ are shown for 4 contacts (left) and 10 contacts (right). We performed rigid body docking on the 6/21 RNA-protein complexes with at least 75% true positives for the top 4 contacts. (B) The i-RMSD between docked models and the experimental structure is plotted for docking runs with EC-derived distance restraints (red) and center-of-mass restraints only (gray). The inclusion of EC-derived distance restraints leads to a significant decrease in i-RMSD.

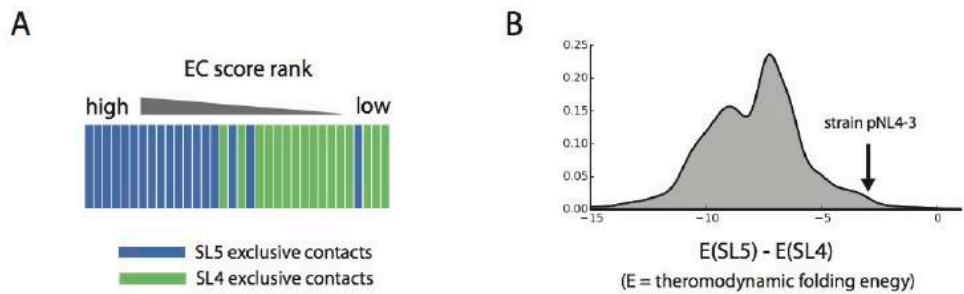

**Figure S6: Additional data for HIV Rev Response Element (related to Figure 6)**
The plots here follow an identical analysis as those shown in Figure 7, but using the LANL alignment of RRE, as opposed to RFAM. Briefly, if we define contacts as SL4- or SL5- exclusive if they satisfy one of these RRE secondary structures but not the other, then the SL5- exclusive contacts largely outrank the SL4- exclusive contacts (A). Studies reporting the SL4 structure have mostly used the pNL4-3 variant of HIV, which is an outlier for favoring SL4 (B) according to thermodynamic folding energy predictions computed on the LANL alignment.

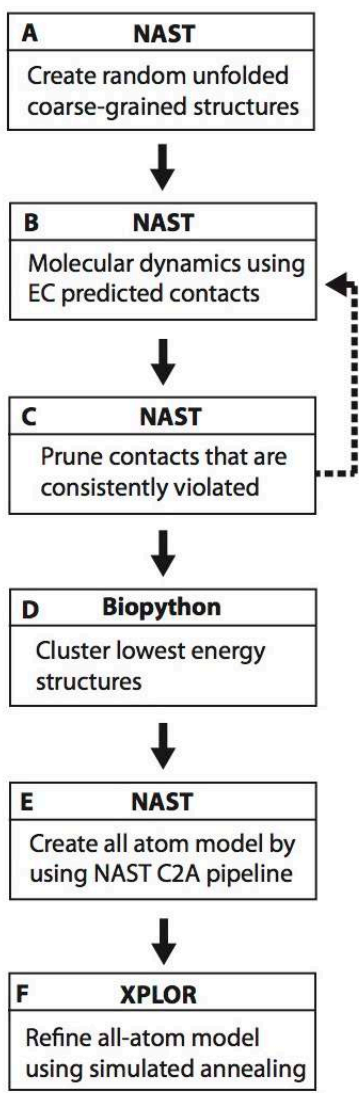

**Figure S7: RNA 3D structure prediction pipeline (related to Exp. procedures)**
We performed coarse-grained modeling in NAST, generating random secondary structures (A) that were then folded with ECs (B). We removed consistently violated contacts with an iterative pruning procedure (C) and clustered the resulting decoys (D) for conversion to all-atom structures (E) and refinement in XPLOR (F).

# References

Adachi, A., Gendelman, H.E., Koenig, S., Folks, T., Willey, R., Rabson, A., and Martin, M.A. (1986). Production of acquired immunodeficiency syndrome-associated retrovirus in human and nonhuman cells transfected with an infectious molecular clone. J Virol *59*, 284-291.

Bai, Y., Tambe, A., Zhou, K., and Doudna, J.A. (2014). RNA-guided assembly of Rev-RRE nuclear export complexes. Elife *3*, e03656.

Bartel, D.P., Zapp, M.L., Green, M.R., and Szostak, J.W. (1991). HIV-1 Rev regulation involves recognition of non-Watson-Crick base pairs in viral RNA. Cell *67*, 529-536.

Berman, H.M., Westbrook, J., Feng, Z., Gilliland, G., Bhat, T.N., Weissig, H., Shindyalov, I.N., and Bourne, P.E. (2000). The Protein Data Bank. Nucleic Acids Res *28*, 235-242.

Besag, J. (1975). Statistical analysis of non-lattice data. The statistician, 179-195.

Burge, S.W., Daub, J., Eberhardt, R., Tate, J., Barquist, L., Nawrocki, E.P., Eddy, S.R., Gardner, P.P., and Bateman, A. (2013). Rfam 11.0: 10 years of RNA families. Nucleic Acids Res *41*, D226-232.

Butcher, S.E., and Pyle, A.M. (2011). The molecular interactions that stabilize RNA tertiary structure: RNA motifs, patterns, and networks. Acc Chem Res *44*, 1302-1311.

Cao, S., and Chen, S.-J. (2011). Physics-based de novo prediction of RNA 3D structures. The journal of physical chemistry B *115*, 4216-4226.

Caserta, E., Liu, L.C., Grundy, F.J., and Henkin, T.M. (2015). Codon-Anticodon Recognition in the Bacillus subtilis glyQS T Box Riboswitch: RNA-DEPENDENT CODON SELECTION OUTSIDE THE RIBOSOME. J Biol Chem *290*, 23336-23347.

Casu, F., Duggan, B.M., and Hennig, M. (2013). The arginine-rich RNA-binding motif of HIV-1 Rev is intrinsically disordered and folds upon RRE binding. Biophys J *105*, 1004-1017.

Charpentier, B., Stutz, F., and Rosbash, M. (1997). A dynamic in vivo view of the HIV-I Rev-RRE interaction. J Mol Biol *266*, 950-962.

Cheng, C.Y., Chou, F.C., Kladwang, W., Tian, S., Cordero, P., and Das, R. (2015). Consistent global structures of complex RNA states through multidimensional chemical mapping. Elife *4*, e07600.

Crothers, D.M., Seno, T., and Soll, G. (1972). Is there a discriminator site in transfer RNA? Proc Natl Acad Sci U S A *69*, 3063-3067.

Das, R., and Baker, D. (2007). Automated de novo prediction of native-like RNA tertiary structures. Proc Natl Acad Sci U S A *104*, 14664-14669.

Das, R., Karanicolas, J., and Baker, D. (2010). Atomic accuracy in predicting and designing noncanonical RNA structure. Nat Methods *7*, 291-294.

Daugherty, M.D., Booth, D.S., Jayaraman, B., Cheng, Y., and Frankel, A.D. (2010). HIV Rev response element (RRE) directs assembly of the Rev homooligomer into discrete asymmetric complexes. Proc Natl Acad Sci U S A *107*, 12481-12486.

Davis, I.W., Murray, L.W., Richardson, J.S., and Richardson, D.C. (2004). MOLPROBITY: structure validation and all-atom contact analysis for nucleic acids and their complexes. Nucleic Acids Res *32*, W615-619.

Ding, Y., Tang, Y., Kwok, C.K., Zhang, Y., Bevilacqua, P.C., and Assmann, S.M. (2014). In vivo genome-wide profiling of RNA secondary structure reveals novel regulatory features. Nature *505*, 696-700.

Dominguez, C., Boelens, R., and Bonvin, A.M. (2003). HADDOCK: a protein-protein docking approach based on biochemical or biophysical information. J Am Chem Soc *125*, 1731-1737.

Dunn, S.D., Wahl, L.M., and Gloor, G.B. (2008). Mutual information without the influence of phylogeny or entropy dramatically improves residue contact prediction. Bioinformatics *24*, 333-340.

Dutheil, J.Y., Jossinet, F., and Westhof, E. (2010). Base pairing constraints drive structural epistasis in ribosomal RNA sequences. Mol Biol Evol *27*, 1868-1876.

Eddy, S.R. (1998). Profile hidden Markov models. Bioinformatics *14*, 755-763.

Eddy, S.R. (2014). Computational analysis of conserved RNA secondary structure in transcriptomes and genomes. Annu Rev Biophys *43*, 433-456.

Ehresmann, C., Baudin, F., Mougel, M., Romby, P., Ebel, J.P., and Ehresmann, B. (1987). Probing the structure of RNAs in solution. Nucleic Acids Res *15*, 9109-9128.

Ekeberg, M., Lovkvist, C., Lan, Y., Weigt, M., and Aurell, E. (2013). Improved contact prediction in proteins: using pseudolikelihoods to infer Potts models. Phys Rev E Stat Nonlin Soft Matter Phys *87*, 012707.

Evans, D., Marquez, S.M., and Pace, N.R. (2006). RNase P: interface of the RNA and protein worlds. Trends Biochem Sci *31*, 333-341.

Fernandes, J., Jayaraman, B., and Frankel, A. (2012). The HIV-1 Rev response element: an RNA scaffold that directs the cooperative assembly of a homo-oligomeric ribonucleoprotein complex. RNA Biol *9*, 6-11.

Finn, R.D., Bateman, A., Clements, J., Coggill, P., Eberhardt, R.Y., Eddy, S.R., Heger, A., Hetherington, K., Holm, L., Mistry, J.*, et al.* (2014). Pfam: the protein families database. Nucleic Acids Res *42*, D222-230.

Finn, R.D., Clements, J., and Eddy, S.R. (2011). HMMER web server: interactive sequence similarity searching. Nucleic Acids Res *39*, W29-37.

Fox, G.E., and Woese, C.R. (1975). 5S RNA secondary structure. Nature *256*, 505-507.

Frellsen, J., Moltke, I., Thiim, M., Mardia, K.V., Ferkinghoff-Borg, J., and Hamelryck, T. (2009). A probabilistic model of RNA conformational space. PLoS Comput Biol *5*, e1000406.

Freyhult, E., Moulton, V., and Gardner, P. (2005). Predicting RNA structure using mutual information. Appl Bioinformatics *4*, 53-59.

Gallego, J., and Varani, G. (2001). Targeting RNA with small-molecule drugs: therapeutic promise and chemical challenges. Acc Chem Res *34*, 836-843.

Garneau, N.L., Wilusz, J., and Wilusz, C.J. (2007). The highways and byways of mRNA decay. Nat Rev Mol Cell Biol *8*, 113-126.

Garst, A.D., Edwards, A.L., and Batey, R.T. (2011). Riboswitches: structures and mechanisms. Cold Spring Harb Perspect Biol *3*.

Gopalan, V. (2007). Uniformity amid diversity in RNase P. Proc Natl Acad Sci U S A *104*, 2031-2032.

Green, N.J., Grundy, F.J., and Henkin, T.M. (2010). The T box mechanism: tRNA as a regulatory molecule. FEBS Lett *584*, 318-324.

Grigg, J.C., and Ke, A. (2013). Structural determinants for geometry and information decoding of tRNA by T box leader RNA. Structure *21*, 2025-2032.

Grundy, F.J., Rollins, S.M., and Henkin, T.M. (1994). Interaction between the acceptor end of tRNA and the T box stimulates antitermination in the Bacillus subtilis tyrS gene: a new role for the discriminator base. J Bacteriol *176*, 4518-4526.

Grundy, F.J., Winkler, W.C., and Henkin, T.M. (2002). tRNA-mediated transcription antitermination in vitro: codon-anticodon pairing independent of the ribosome. Proc Natl Acad Sci U S A *99*, 11121-11126.

Gutell, R.R., Power, A., Hertz, G.Z., Putz, E.J., and Stormo, G.D. (1992). Identifying constraints on the higher-order structure of RNA: continued development and application of comparative sequence analysis methods. Nucleic Acids Res *20*, 5785-5795.

Heaphy, S., Finch, J.T., Gait, M.J., Karn, J., and Singh, M. (1991). Human immunodeficiency virus type 1 regulator of virion expression, rev, forms nucleoprotein filaments after binding to a purine-rich "bubble" located within the rev-responsive region of viral mRNAs. Proc Natl Acad Sci U S A *88*, 7366-7370.

Hofacker, I.L., Fekete, M., and Stadler, P.F. (2002). Secondary structure prediction for aligned RNA sequences. J Mol Biol *319*, 1059-1066.

Hofacker, I.L., and Lorenz, R. (2014). Predicting RNA structure: advances and limitations. Methods Mol Biol *1086*, 1-19.

Hopf, T.A., Colwell, L.J., Sheridan, R., Rost, B., Sander, C., and Marks, D.S. (2012). Three-dimensional structures of membrane proteins from genomic sequencing. Cell *149*, 1607-1621.

Hopf, T.A., Ingraham, J.B., Poelwijk, F.J., Springer, M., Sander, C., and Marks, D.S. (2015). Quantification of the effect of mutations using a global probability model of natural sequence variation. arXiv preprint arXiv:151004612.

Hopf, T.A., Scharfe, C.P., Rodrigues, J.P., Green, A.G., Kohlbacher, O., Sander, C., Bonvin, A.M., and Marks, D.S. (2014). Sequence co-evolution gives 3D contacts and structures of protein complexes. Elife *3*.

Huang, W., Thomas, B., Flynn, R.A., Gavzy, S.J., Wu, L., Kim, S.V., Hall, J.A., Miraldi, E.R., Ng, C.P., Rigo, F.W.*, et al.* (2015). DDX5 and its associated lncRNA Rmrp modulate TH17 cell effector functions. Nature *528*, 517-522.

Ippolito, J.A., and Steitz, T.A. (2000). The structure of the HIV-1 RRE high affinity rev binding site at 1.6 A resolution. J Mol Biol *295*, 711-717.

Iwai, S., Pritchard, C., Mann, D.A., Karn, J., and Gait, M.J. (1992). Recognition of the high affinity binding site in rev-response element RNA by the human immunodeficiency virus type-1 rev protein. Nucleic Acids Res *20*, 6465-6472.

Jonikas, M.A., Radmer, R.J., Laederach, A., Das, R., Pearlman, S., Herschlag, D., and Altman, R.B. (2009). Coarse-grained modeling of large RNA molecules with knowledge-based potentials and structural filters. RNA *15*, 189-199.

Kamisetty, H., Ovchinnikov, S., and Baker, D. (2013). Assessing the utility of coevolution-based residue-residue contact predictions in a sequence- and structure-rich era. Proc Natl Acad Sci U S A *110*, 15674-15679.

Khanova, E., Esakova, O., Perederina, A., Berezin, I., and Krasilnikov, A.S. (2012). Structural organizations of yeast RNase P and RNase MRP holoenzymes as revealed by UV-crosslinking studies of RNA-protein interactions. RNA *18*, 720-728.

Kjems, J., Calnan, B.J., Frankel, A.D., and Sharp, P.A. (1992). Specific binding of a basic peptide from HIV-1 Rev. EMBO J *11*, 1119-1129.

Klingler, T.M., and Brutlag, D.L. (1993). Detection of correlations in tRNA sequences with structural implications. Proc Int Conf Intell Syst Mol Biol *1*, 225-233.

Krasilnikov, A.S., Xiao, Y., Pan, T., and Mondragon, A. (2004). Basis for structural diversity in homologous RNAs. Science *306*, 104-107.

Laing, C., and Schlick, T. (2010). Computational approaches to 3D modeling of RNA. J Phys Condens Matter *22*, 283101.

Latham, J.A., and Cech, T.R. (1989). Defining the inside and outside of a catalytic RNA molecule. Science *245*, 276-282.

Lee, C.P., Mandal, N., Dyson, M.R., and RajBhandary, U.L. (1993). The discriminator base influences tRNA structure at the end of the acceptor stem and possibly its interaction with proteins. Proc Natl Acad Sci U S A *90*, 7149-7152.

Legiewicz, M., Badorrek, C.S., Turner, K.B., Fabris, D., Hamm, T.E., Rekosh, D., Hammarskjold, M.L., and Le Grice, S.F. (2008). Resistance to RevM10 inhibition reflects a conformational switch in the HIV-1 Rev response element. Proc Natl Acad Sci U S A *105*, 14365-14370.

Lescoute, A., Leontis, N.B., Massire, C., and Westhof, E. (2005). Recurrent structural RNA motifs, Isostericity Matrices and sequence alignments. Nucleic Acids Res *33*, 2395-2409.

Levitt, M. (1969). Detailed molecular model for transfer ribonucleic acid. Nature *224*, 759-763.

Lorenz, R., Bernhart, S.H., Honer Zu Siederdissen, C., Tafer, H., Flamm, C., Stadler, P.F., and Hofacker, I.L. (2011). ViennaRNA Package 2.0. Algorithms Mol Biol *6*, 26.

Lu, C., Ding, F., Chowdhury, A., Pradhan, V., Tomsic, J., Holmes, W.M., Henkin, T.M., and Ke, A. (2010). SAM recognition and conformational switching mechanism in the Bacillus subtilis yitJ S box/SAM-I riboswitch. J Mol Biol *404*, 803-818.

Luedtke, N.W., and Tor, Y. (2003). Fluorescence-based methods for evaluating the RNA affinity and specificity of HIV-1 Rev-RRE inhibitors. Biopolymers *70*, 103-119.

Magnus, M., Matelska, D., Lach, G., Chojnowski, G., Boniecki, M.J., Purta, E., Dawson, W., Dunin-Horkawicz, S., and Bujnicki, J.M. (2014). Computational modeling of RNA 3D structures, with the aid of experimental restraints. RNA Biol *11*, 522-536.

Malim, M.H., and Cullen, B.R. (1991). HIV-1 structural gene expression requires the binding of multiple Rev monomers to the viral RRE: implications for HIV-1 latency. Cell *65*, 241-248.

Mann, D.A., Mikaelian, I., Zemmel, R.W., Green, S.M., Lowe, A.D., Kimura, T., Singh, M., Butler, P.J., Gait, M.J., and Karn, J. (1994). A molecular rheostat. Co-operative rev binding to stem I of the rev-response element modulates human immunodeficiency virus type-1 late gene expression. J Mol Biol *241*, 193-207.

Marks, D.S., Colwell, L.J., Sheridan, R., Hopf, T.A., Pagnani, A., Zecchina, R., and Sander, C. (2011). Protein 3D structure computed from evolutionary sequence variation. PLoS One *6*, e28766.

Marks, D.S., Hopf, T.A., and Sander, C. (2012). Protein structure prediction from sequence variation. Nat Biotechnol *30*, 1072-1080.

Martin, K.C., and Ephrussi, A. (2009). mRNA localization: gene expression in the spatial dimension. Cell *136*, 719-730.

Mathews, D.H., Disney, M.D., Childs, J.L., Schroeder, S.J., Zuker, M., and Turner, D.H. (2004). Incorporating chemical modification constraints into a dynamic programming algorithm for prediction of RNA secondary structure. Proc Natl Acad Sci U S A *101*, 7287-7292.

McManus, C.J., and Graveley, B.R. (2011). RNA structure and the mechanisms of alternative splicing. Curr Opin Genet Dev *21*, 373-379.
Miao, Z., Adamiak, R.W., Blanchet, M.F., Boniecki, M., Bujnicki, J.M., Chen, S.J., Cheng, C., Chojnowski, G., Chou, F.C., Cordero, P*., et al.* (2015). RNA-Puzzles Round II: assessment of RNA structure prediction programs applied to three large RNA structures. RNA *21*, 1066-1084.
Michel, F., and Westhof, E. (1990). Modelling of the three-dimensional architecture of group I catalytic introns based on comparative sequence analysis. J Mol Biol *216*, 585-610.
Moazed, D., and Noller, H.F. (1986). Transfer RNA shields specific nucleotides in 16S ribosomal RNA from attack by chemical probes. Cell *47*, 985-994.
Mokdad, A., and Frankel, A.D. (2008). ISFOLD: structure prediction of base pairs in non-helical RNA motifs from isostericity signatures in their sequence alignments. J Biomol Struct Dyn *25*, 467-472.
Morcos, F., Pagnani, A., Lunt, B., Bertolino, A., Marks, D.S., Sander, C., Zecchina, R., Onuchic, J.N., Hwa, T., and Weigt, M. (2011). Direct-coupling analysis of residue coevolution captures native contacts across many protein families. Proc Natl Acad Sci U S A *108*, E1293-1301.
Mortimer, S.A., Kidwell, M.A., and Doudna, J.A. (2014). Insights into RNA structure and function from genome-wide studies. Nat Rev Genet *15*, 469-479.
Nawrocki, E.P., Burge, S.W., Bateman, A., Daub, J., Eberhardt, R.Y., Eddy, S.R., Floden, E.W., Gardner, P.P., Jones, T.A., Tate, J*., et al.* (2015). Rfam 12.0: updates to the RNA families database. Nucleic Acids Res *43*, D130-137.
Nawrocki, E.P., and Eddy, S.R. (2013). Infernal 1.1: 100-fold faster RNA homology searches. Bioinformatics *29*, 2933-2935.
Novikova, I.V., Hennelly, S.P., and Sanbonmatsu, K.Y. (2012). Sizing up long non-coding RNAs: do lncRNAs have secondary and tertiary structure? Bioarchitecture *2*, 189-199.
Nussinov, R., and Jacobson, A.B. (1980). Fast algorithm for predicting the secondary structure of single-stranded RNA. Proc Natl Acad Sci U S A *77*, 6309-6313.
Olsen, H.S., Nelbock, P., Cochrane, A.W., and Rosen, C.A. (1990). Secondary structure is the major determinant for interaction of HIV rev protein with RNA. Science *247*, 845-848.
Ovchinnikov, S., Kamisetty, H., and Baker, D. (2014). Robust and accurate prediction of residue-residue interactions across protein interfaces using evolutionary information. Elife *3*, e02030.
Pang, P.S., Jankowsky, E., Wadley, L.M., and Pyle, A.M. (2005). Prediction of functional tertiary interactions and intermolecular interfaces from primary sequence data. J Exp Zool B Mol Dev Evol *304*, 50-63.
Parisien, M., and Major, F. (2008). The MC-Fold and MC-Sym pipeline infers RNA structure from sequence data. Nature *452*, 51-55.
Peterson, R.D., and Feigon, J. (1996). Structural change in Rev responsive element RNA of HIV-1 on binding Rev peptide. J Mol Biol *264*, 863-877.
Petrov, A.I., Zirbel, C.L., and Leontis, N.B. (2011). WebFR3D--a server for finding, aligning and analyzing recurrent RNA 3D motifs. Nucleic Acids Res *39*, W50-55.
Pollom, E., Dang, K.K., Potter, E.L., Gorelick, R.J., Burch, C.L., Weeks, K.M., and Swanstrom, R. (2013). Comparison of SIV and HIV-1 genomic RNA structures reveals impact of sequence evolution on conserved and non-conserved structural motifs. PLoS Pathog *9*, e1003294.
Quinodoz, S., and Guttman, M. (2014). Long noncoding RNAs: an emerging link between gene regulation and nuclear organization. Trends Cell Biol *24*, 651-663.

Ramani, V., Qiu, R., and Shendure, J. (2015). High-throughput determination of RNA structure by proximity ligation. Nat Biotechnol.

Rausch, J.W., and Le Grice, S.F. (2015). HIV Rev Assembly on the Rev Response Element (RRE): A Structural Perspective. Viruses *7*, 3053-3075.

Reiter, N.J., Osterman, A., Torres-Larios, A., Swinger, K.K., Pan, T., and Mondragon, A. (2010). Structure of a bacterial ribonuclease P holoenzyme in complex with tRNA. Nature *468*, 784-789.

Rinn, J.L., and Chang, H.Y. (2012). Genome regulation by long noncoding RNAs. Annu Rev Biochem *81*, 145-166.

Rivas, E., and Eddy, S.R. (1999). A dynamic programming algorithm for RNA structure prediction including pseudoknots. J Mol Biol *285*, 2053-2068.

Rouskin, S., Zubradt, M., Washietl, S., Kellis, M., and Weissman, J.S. (2014). Genome-wide probing of RNA structure reveals active unfolding of mRNA structures in vivo. Nature *505*, 701-705.

Rutherford, S.T., Valastyan, J.S., Taillefumier, T., Wingreen, N.S., and Bassler, B.L. (2015). Comprehensive analysis reveals how single nucleotides contribute to noncoding RNA function in bacterial quorum sensing. Proc Natl Acad Sci U S A *112*, E6038-6047.

Sarver, M., Zirbel, C.L., Stombaugh, J., Mokdad, A., and Leontis, N.B. (2008). FR3D: finding local and composite recurrent structural motifs in RNA 3D structures. J Math Biol *56*, 215-252.

Schwieters, C.D., Kuszewski, J.J., Tjandra, N., and Clore, G.M. (2003). The Xplor-NIH NMR molecular structure determination package. J Magn Reson *160*, 65-73.

Serganov, A., and Patel, D.J. (2012). Molecular recognition and function of riboswitches. Curr Opin Struct Biol *22*, 279-286.

Serganov, A., Polonskaia, A., Phan, A.T., Breaker, R.R., and Patel, D.J. (2006). Structural basis for gene regulation by a thiamine pyrophosphate-sensing riboswitch. Nature *441*, 1167-1171.

Shang, L., Xu, W., Ozer, S., and Gutell, R.R. (2012). Structural constraints identified with covariation analysis in ribosomal RNA. PLoS One *7*, e39383.

Sherpa, C., Rausch, J.W., Le Grice, S.F., Hammarskjold, M.L., and Rekosh, D. (2015). The HIV-1 Rev response element (RRE) adopts alternative conformations that promote different rates of virus replication. Nucleic Acids Res *43*, 4676-4686.

Sigova, A.A., Abraham, B.J., Ji, X., Molinie, B., Hannett, N.M., Guo, Y.E., Jangi, M., Giallourakis, C.C., Sharp, P.A., and Young, R.A. (2015). Transcription factor trapping by RNA in gene regulatory elements. Science *350*, 978-981.

Spitale, R.C., Flynn, R.A., Zhang, Q.C., Crisalli, P., Lee, B., Jung, J.W., Kuchelmeister, H.Y., Batista, P.J., Torre, E.A., Kool, E.T.*, et al.* (2015). Structural imprints in vivo decode RNA regulatory mechanisms. Nature *519*, 486-490.

Sreedhara, A., and Cowan, J.A. (2001). Targeted site-specific cleavage of HIV-1 viral Rev responsive element by copper aminoglycosides. J Biol Inorg Chem *6*, 166-172.

Turner, D.H., and Mathews, D.H. (2010). NNDB: the nearest neighbor parameter database for predicting stability of nucleic acid secondary structure. Nucleic Acids Res *38*, D280-282.

UniProt, C. (2015). UniProt: a hub for protein information. Nucleic Acids Res *43*, D204-212.

Wan, Y., Qu, K., Zhang, Q.C., Flynn, R.A., Manor, O., Ouyang, Z., Zhang, J., Spitale, R.C., Snyder, M.P., Segal, E.*, et al.* (2014). Landscape and variation of RNA secondary structure across the human transcriptome. Nature *505*, 706-709.

Warf, M.B., and Berglund, J.A. (2010). Role of RNA structure in regulating pre-mRNA splicing. Trends in Biochemical Sciences *35*, 169-178.

Weigt, M., White, R.A., Szurmant, H., Hoch, J.A., and Hwa, T. (2009). Identification of direct residue contacts in protein-protein interaction by message passing. Proc Natl Acad Sci U S A *106*, 67-72.

Zemmel, R.W., Kelley, A.C., Karn, J., and Butler, P.J. (1996). Flexible regions of RNA structure facilitate co-operative Rev assembly on the Rev-response element. J Mol Biol *258*, 763-777.

Zhang, J., and Ferre-D'Amare, A.R. (2013). Co-crystal structure of a T-box riboswitch stem I domain in complex with its cognate tRNA. Nature *500*, 363-366.

Zuker, M. (2003). Mfold web server for nucleic acid folding and hybridization prediction. Nucleic Acids Res *31*, 3406-3415.